**Towards In-Situ Failure Assessment: Deep Learning on DIC Results for Laminated Composites**

Amir Mohammad Mirzaei[1]

*Faculty of Engineering and Applied Science, Cranfield University, Cranfield MK43 0AL, UK*

## Abstract

Predicting fracture load in laminated composites with stress raisers is challenging due to complex failure mechanisms such as delamination, fibre breakage, and matrix cracking, which are heavily influenced by fibre orientation, layup sequence, and notch geometry. This study aims to address this by developing a novel deep learning framework that leverages solely experimental strain field data from Digital Image Correlation (DIC) for accurate, in-situ predictions—bypassing the need for finite element simulations or empirical calibrations. Two complementary architectures are explored: a multi-layer perceptron (MLP) that processes numerical values of maximum principal strain from a targeted rectangular region ahead of the notch, enhanced by advanced feature selection (mutual information, Lasso, and SHAP) to focus on critical data points; and a convolutional neural network (CNN) trained on full-field strain images, bolstered by data augmentation to handle variability and prevent overfitting. Validated across 116 quasi-static tests encompassing 31 distinct configurations—including six layups (quasi-isotropic to highly anisotropic) with four off-axis angles for open-hole specimens, and one cross-ply layup with four off-axis and four on-axis notch orientations for U-notched specimens—the MLP and CNN achieve coefficients of determination ($R^2$) of 0.86 and 0.82, respectively. This framework captures a broad spectrum of damage modes and responses, from brittle fibre-dominated fracture to ductile delamination-driven failure, and due to its computational efficiency and reliance only on DIC measurements, the approach enables practical in-situ fracture load estimation.

[1]Corresponding author. Email address: amir.mirzaei@cranfield.ac.uk

**Keywords:** Laminated Composites; Notch; Digital Image Correlation; Machine Learning; Artificial Neural Networks

## 1. Introduction

The anisotropic behaviour of laminated composites poses significant challenges in predicting fracture loads for notched structures, where intricate interactions among failure mechanisms— such as fibre breakage, matrix cracking, and delamination— often prevent straightforward analysis through conventional models, particularly under the influence of variables like fibre orientation, stacking sequence, and notch shape [1,2]. Classical approaches such as the point stress/average stress criteria by Whitney and Nuismer [3] were among the first to relate stress distributions around notches to laminate failure strengths. These methods introduced a characteristic length over which the stress is averaged to account for damage tolerance, but the optimal parameters often depend on empirical calibration. Another early contribution to understanding notch sensitivity in composites was made by Pipes et al. [4]. They proposed a notch strength model that accounted for the influence of notch size and material anisotropy on the failure of laminated composites, providing a complementary perspective to stress-based criteria. In contrast, modern full-field techniques like digital image correlation (DIC) [5] have revolutionized damage visualization, allowing precise tracking of strain patterns and evolving degradation in composites [6]. Initial experiments underscored how pre-catastrophic damage, including matrix cracks and delaminations, reshapes strain fields and diminishes residual load capacity in notched laminates [7]. For instance, Ambu et al. [8] leveraged DIC to trace progressive failure in notched carbon/epoxy and carbon/PEEK systems, revealing how varied mechanisms—fibre rupture, delamination, and beyond—trigger unique strain redistributions that directly impact strength. Such DIC insights have further illuminated the dual-edged role of subcritical damage in modulating stress raisers, either amplifying or mitigating them to influence ultimate laminate endurance, positioning DIC as an instrument for dissecting failure dynamics and their effects on structural integrity [9]. With advances in computational power, researchers have increasingly turned to machine learning (ML) to use strain field data—both from experiments (e.g., DIC) and high-fidelity simulations—in order to predict fracture loads more accurately across different notch geometries, off-axis fibre orientations, and laminate layups [10].

Given the complexity of composite failure and the richness of DIC data, machine learning offers a promising approach to model and predict fracture loads more accurately. Azeem and Pinho [11] introduced a machine-learning-enhanced characteristic length method for open-hole tension laminates. In their approach, multiple regression models were trained to directly predict the ultimate failure load of an open-hole composite specimen. By learning from a database of finite element simulations and experiments, their ML models could capture the influence of variables like hole size and plate width that go beyond the ideal infinite-width assumption of analytical solutions, yielding more accurate strength predictions than the base point-stress model. Balducci et al. [12] redefined open-hole failure prediction as a classification problem, learning the failure envelope (the set of stress combinations leading to failure) for a composite plate with a hole. They trained Support Vector Machine classifiers to distinguish safe vs. failure loading states and showed that their best SVM (with a tuned radial-basis kernel) achieved over 90% accuracy in classifying failure vs. non-failure cases. This demonstrates that even "traditional" ML algorithms (SVMs, decision trees, etc.), when provided with sufficient training data from simulations or experiments, can learn complex failure criteria in composites. Beyond open-hole tension, researchers have applied ML to other loading and geometric scenarios. Li et al. [13] proposed a high-throughput simulation and ML framework to predict the allowable compression load of notched laminates under compression. They generated a large dataset of finite element analyses (FEA) for various notch sizes, laminate layups, and loading conditions, and then trained regression models to map these input parameters to the laminate's critical buckling or failure load. This approach effectively produced a surrogate model that can instantly predict the notched compression strength given a set of design parameters, eliminating the need for exhaustive incremental testing. Similarly, a deep transfer learning approach by Li et al. [14] constructed an entire allowable load space for notched composites under different design configurations. In that study, an ensemble of deep neural networks was first trained on a broad range of notch geometries, materials (fibre and matrix properties), stacking sequences (including various off-axis ply angles and ply counts), and loading types (tension, compression, biaxial). This ensemble DNN could predict the failure load for a given set of design parameters within the training distribution.

While the above approaches often use summary features (like geometric parameters or aggregate stress values), an alternative paradigm is to employ the full-field strain data from experiments or high-fidelity simulations as direct input to machine learning models. The idea is that the strain field reflects both stress concentrations and damage states, thus containing rich information for predicting the load-bearing capacity. Nastos et al. [15] demonstrated this concept in a non-destructive strength prediction method for composites. In their work, a quasi-isotropic carbon/epoxy laminate with a central hole was loaded to only 20% of its expected failure load, and the surface strain field was measured with DIC. These moderate-load strain maps – well before any visible damage or cracking – were then fed into a deep learning model (a convolutional neural network) that had been trained on thousands of random FEA simulations of the same laminate under random strength properties. The CNN learned to correlate the pattern of strains at 10–20% load with the eventual ultimate tensile strength of the specimen. When tested on actual specimens (loaded to 10–20% and then unloaded), the model predicted failure loads with good accuracy and low error variability. Another use of full-field strain data is in damage identification and early warning of failure. Rather than directly outputting a numeric failure load, some studies train ML models to detect and locate damage from strain patterns, which can indirectly predict failure. Wang et al. [16] developed a CNN-based semantic segmentation model to analyze DIC strain maps of CFRP laminates with open-holes. Using a DeepLabv3+ architecture [17] (with a ResNet-50 backbone [18]), they trained the network to perform pixel-level classification of the strain field – essentially labeling regions as damaged (plasticity, fibre break, delamination) or healthy based on the strain distribution. The training data were generated via high-fidelity finite element simulations of tension tests, ensuring the network saw a wide variety of damage patterns. When applied to real-time DIC data from composite coupons, the model achieved a high segmentation accuracy (mean IoU ≈ 0.92) in identifying damage zones. This automated interpretation of strain contours means that the onset of critical damage can be detected online, without relying on a human inspector.

Following the reviewed studies, this paper builds on previous investigations by the author to employ machine learning models trained only on field distributions to predict the failure of notched components. In [19], stress, strain, and energy fields were utilized to estimate the fatigue life of notched components. The framework was validated across diverse loading conditions, notch geometries, and materials, including steel, additively manufactured PLA, and laminated composites [20]. The study demonstrated that all field distributions effectively predicted fatigue life, with stress fields generally providing the most stable and accurate predictions across all cases. This framework was further extended in [21] to predict fracture load (defined as maximum load based on load-displacement curve) and crack initiation angle under mixed-mode in-plane loading conditions in cracked bodies. The model was validated using five distinct specimen configurations, each subjected to at least six different mode mixity conditions. Results showed that stress field data, when combined with a multi-layer perceptron (MLP), outperformed conventional models such as Generalized Maximum Tangential Stress (MTS) and Theory of Critical Distances (TCD) in prediction accuracy, and consistency of performance across various input parameter configurations was confirmed.

In this paper, the approach is extended to include not only anisotropic materials but also the direct use of DIC results. This research addresses a critical gap by exclusively utilizing experimental strain field data from Digital Image Correlation (DIC) to train deep learning models for predicting fracture loads across diverse layup configurations and notch geometries, eliminating the need for additional finite element analyses or post-processing. To utilize DIC results, two distinct approaches are employed: first, the use of numerical values of maximum principal strain at representative nodes ahead of the notch; second, images of the principal strain field distribution around the notch obtained from DIC. Studies have shown that the principal strain distribution is sensitive to changes in layup configuration and notch geometry, making it a robust metric for fracture prediction [6,9,16]. For the first approach, an MLP model with stepwise feature selection algorithms is used to handle sparse to dense sets of representative nodes in DIC data. To accommodate different notch geometries, a novel approach is introduced, where nodes are considered within a rectangular region ahead of the notch. For the second approach, a Convolutional Neural Network (CNN) based on the DenseNet-121 architecture is employed. Since both approaches are data-driven and typically require extensive data, the models in this paper are trained with a reasonable number of experimental tests by employing techniques such as feature selection for the MLP and data augmentation for the CNN, enhancing efficiency and applicability across various cases. To this end, no additional finite element analyses are used to generate data, and only strain fields calculated from DIC in the experimental campaign are employed. In Section 2, the experimental campaign is presented. In Section 3, the deep learning algorithms are introduced. In Section 4, the results are presented, and the robustness of the framework across varied input data is demonstrated and discussed.

## 2. Experimental tests

This section presents the experimental data used to validate the proposed framework. In this study, two different geometries are considered: the open-hole tests, whose results were presented by Mitrou et al. [22], and the U-notch specimens, which were manufactured and tested by the same group but are presented in this paper for the first time. The specimens utilized in the experimental campaign were composed of carbon fibre thin-ply carbon/epoxy laminates produced from North Thin-Ply Technology T800/736 LT unidirectional prepreg, with a nominal ply thickness of 0.05 mm and ply properties of $E_1$ = 143.83 GPa, $E_2$ = 7.85 GPa, $G_{12}$ = 3.80 GPa and $\nu_{12}$ = 0.38. Six 32-ply laminates (≈ 1.6 mm thick overall) were tested covering a wide range from quasi-isotropic to highly anisotropic behaviour as defined in [22]: QI $[90/-45/0/45]_{4S}$, SOFT $[90/\pm45/90/\pm45/90/0]_{2S}$, CP30 $[90/0/\pm30/0/90/\pm30]_{2S}$, CP1560 $[90/\pm60/0/\pm15/90/0]_{2S}$, CP1575 $[90/\pm75/0/\pm15/0/90]_{2S}$ and cross-ply CP $[90/0]_{8S}$. All specimens were 20 mm wide and 170 mm long overall, with the effective gauge length of 120 mm. The open-hole specimens had a central hole with a diameter of 5 mm, while the U-notched specimens had a notch 5 mm deep and 1 mm wide, created using a 0.5 mm radius drill bit. It is noted that the U-notched specimens were only manufactured from a CP layup. Off-axis loading was introduced by cutting the coupons in the parent plate at 0°, 15°, 45° and 60° relative to the fibre-0° direction, see Fig. 1a). As stated in [22], each laminate and angle combination had a number of samples that ranged from 3 to 5, giving 109 overall tests for the open-hole configuration. For the U-notched samples, apart from the off-axis testing similar to the open-hole testing, angled notched configuration was introduced, meaning the off-axis angle was fixed and equal to 0°, but the notch direction was in 15°, 45°, 60° and 90° of fibre-0° (loading) direction, which is referred to as the on-axis notch test in this study (Fig. 1b)). For the U-notched specimens only one sample per configuration was tested yielding 7 samples. Notably, the 0° off-axis configuration corresponds to the 90° on-axis case. In total, 116 mechanical tests were conducted across 31 distinct configurations. A comprehensive overview of all tested specimen configurations, including detailed geometrical parameters, layup sequences, and loading conditions, is provided in Table 1.

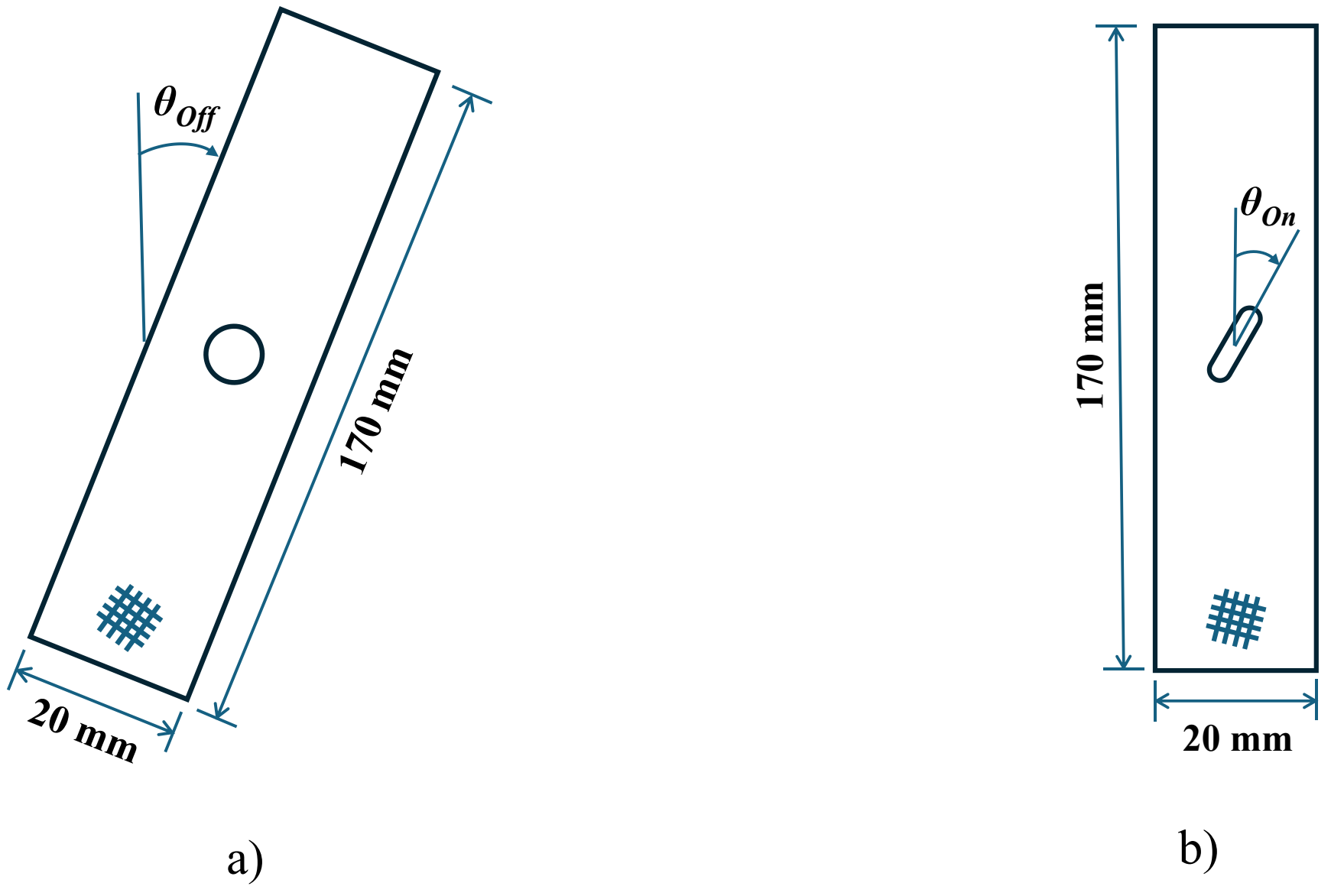

**Fig. 1. Schematic of loading configurations for notched composite specimens. a) Off-axis loading with angles $\theta_{Off}$ = 0°, 15°, 45°, 60° defined relative to the fibre-0° direction** [22]. **b) On-axis loading for U-notched cross-ply (CP) specimens, with angles of $\theta_{On}$ = 15°, 45°, 60°, 90° relative to the loading direction (fibre-0°).**

**Table 1. Summary of specimen geometries, notch dimensions, layup sequences, loading configurations, and the corresponding number of tests for each configuration.**

| Notch Geometry | Notch Dimensions (mm) | Layup | Loading Configuration | Number of Tests |
|---|---|---|---|---|
| Open-hole | Hole radius: 2.5 | QI $[90/\text{-}45/0/45]_{4S}$,<br>SOFT $[90/\pm45/90/\pm45/90/0]_{2S}$,<br>CP30 $[90/0/\pm30/0/90/\pm30]_{2S}$,<br>CP1560 $[90/\pm60/0/\pm15/90/0]_{2S}$,<br>CP1575 $[90/\pm75/0/\pm15/0/90]_{2S}$,<br>CP $[90/0]_{8S}$ | 4 off-axis:<br>$\theta_{Off}$ = 0°, 15°, 45°, 60° | 109 |
| U-notch | Depth: 5<br>Width: 1<br>Root radius: 0.5 | CP $[90/0]_{8S}$ | 4 off-axis:<br>$\theta_{Off}$ = 0°, 15°, 45°, 60°<br>+ 4 on-axis (3 unique):<br>$\theta_{On}$ = 15°, 45°, 60°, 90° | 7 |

To obtain strain fields based on DIC, full-field kinematics of the specimens were monitored with a single-camera 2D digital image correlation setup. Images were processed to extract the strain field in VIC-2D software using a 15-pixel subset, a 7-pixel step, and the fill-boundary option to obtain data close to the notch boundary. In the Appendix, the strain fields obtained using DIC for some arbitrary cases are presented and compared with the finite element results (for validation purposes only).

In order to emphasize and demonstrate the different mechanical responses between configurations, the results of mechanical testing are presented as force–displacement curves for eight different configurations in Fig. 2. In this figure, the subfigure labels indicate the layup configuration, with different colours representing various off-axis angles—except for the U-notched specimens tested under the on-axis configuration (CP_UO), where the angles correspond to the notch direction relative to the fibre-0° (loading) direction.

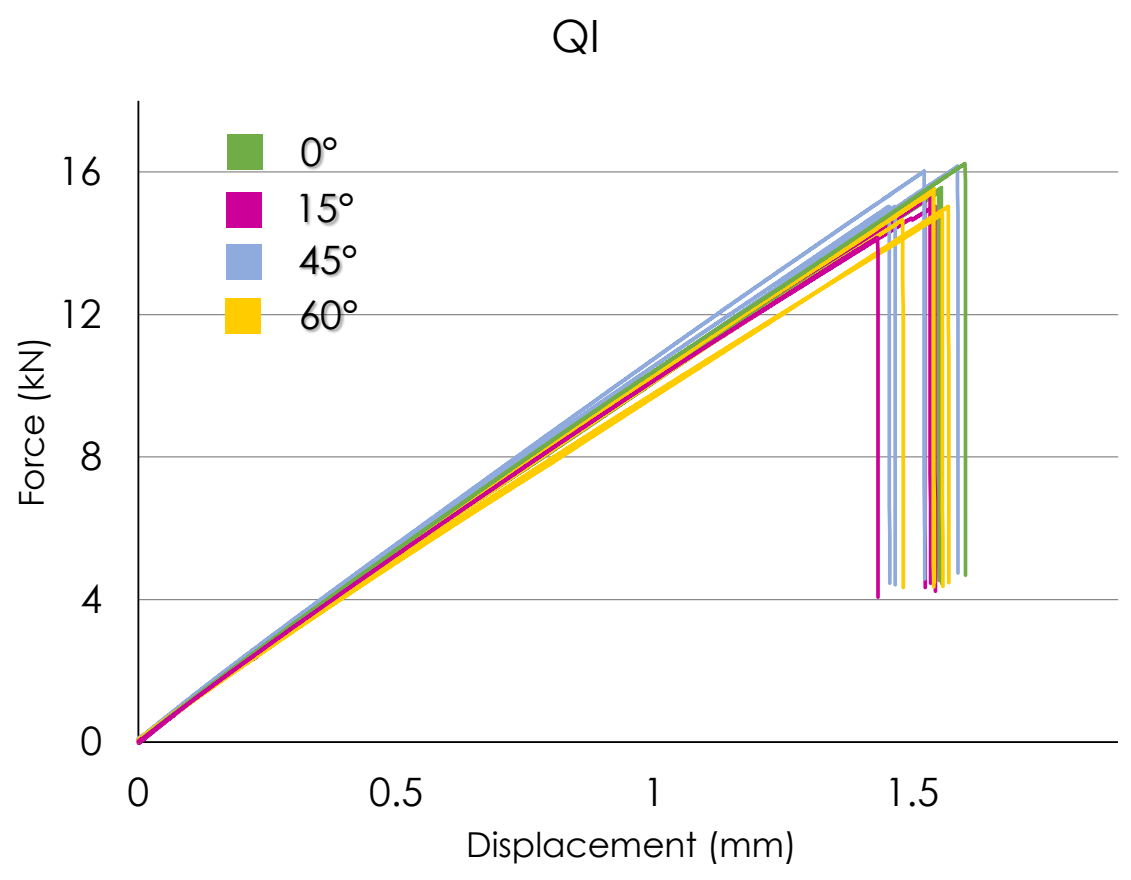

a)

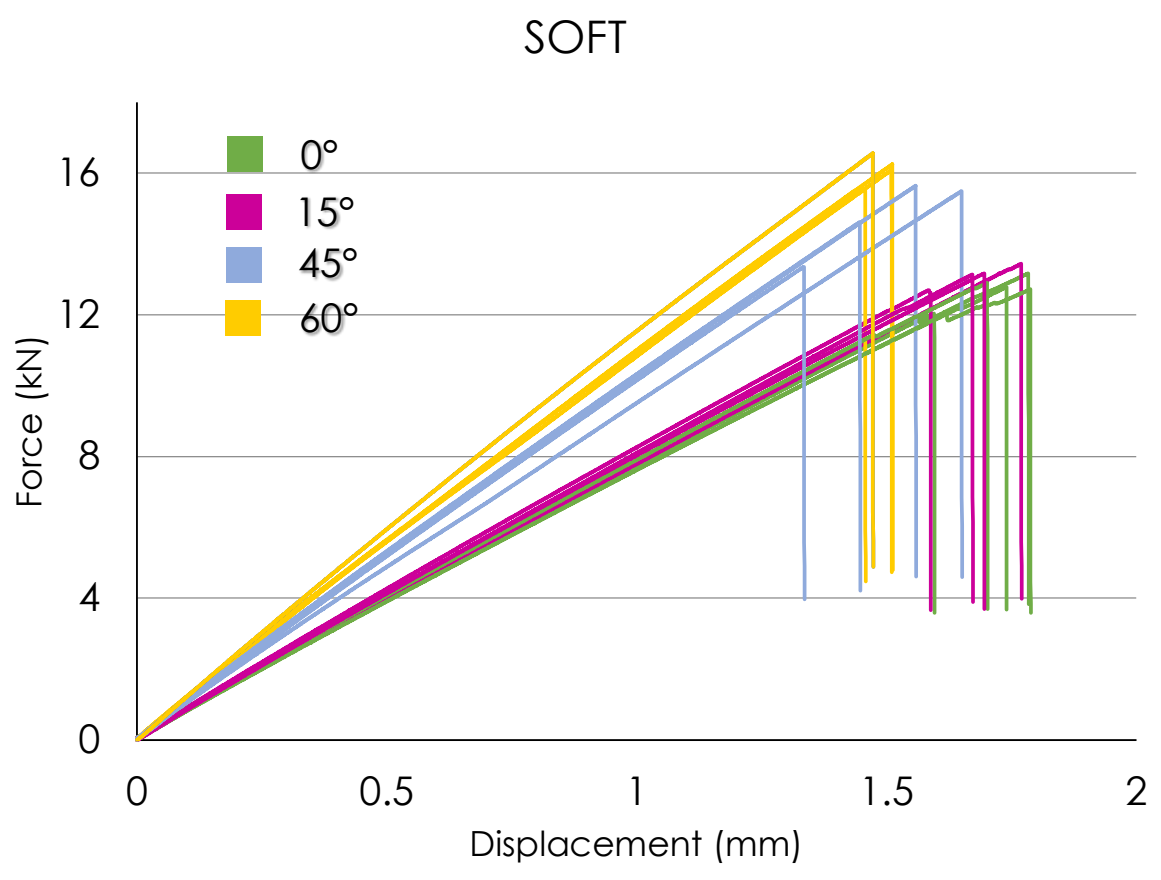

b)

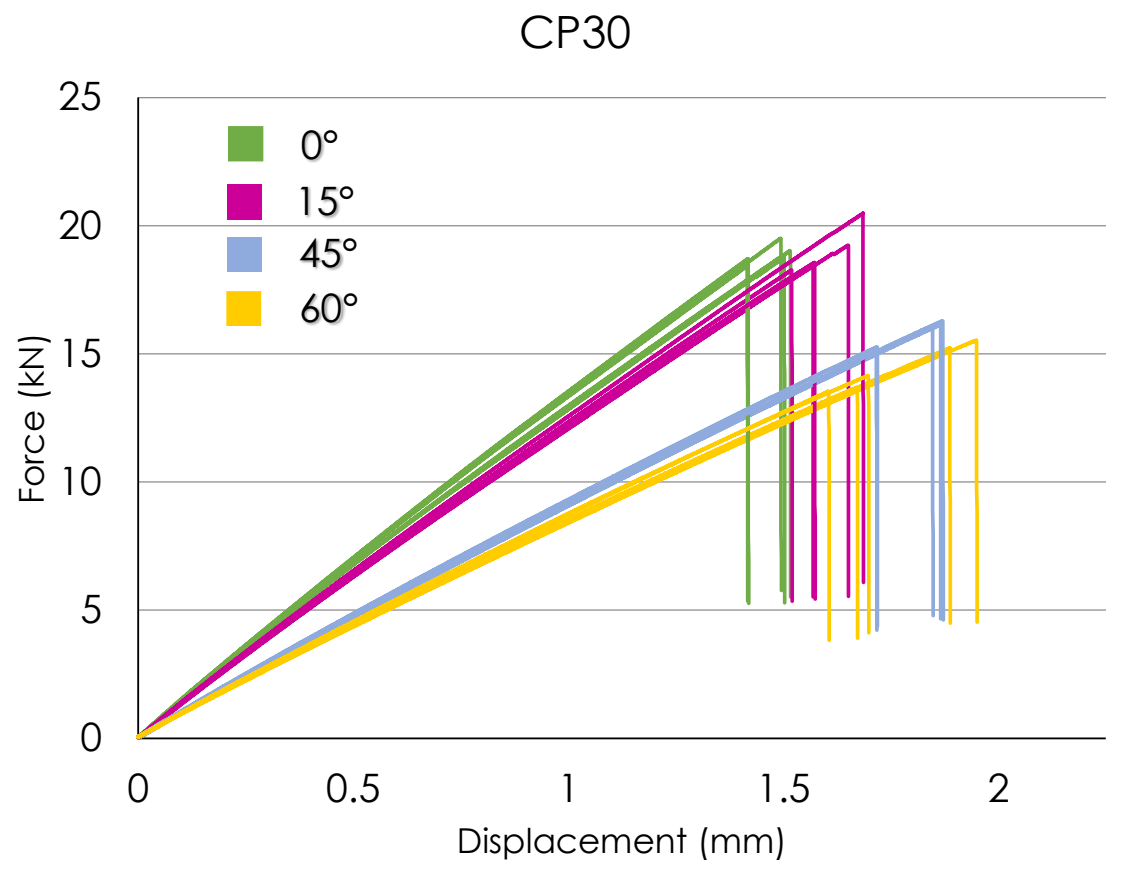

c)

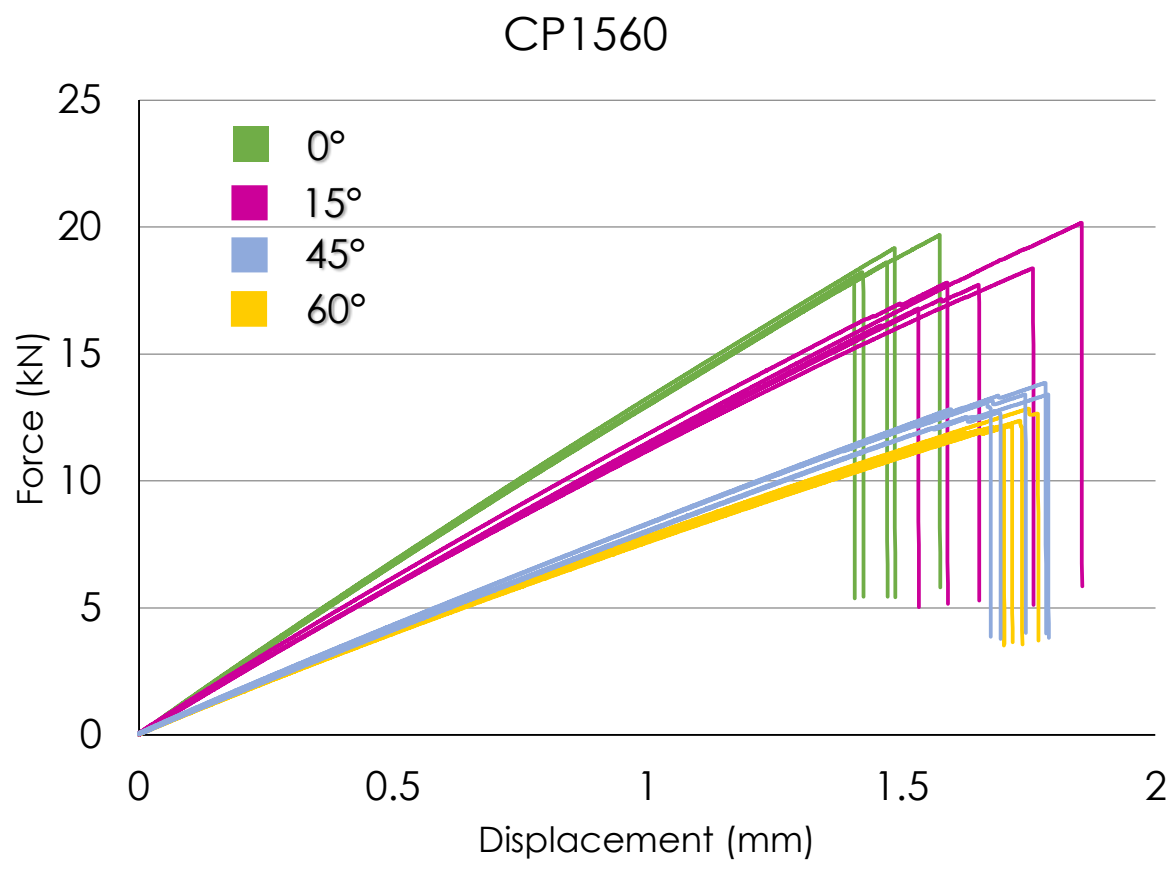

d)

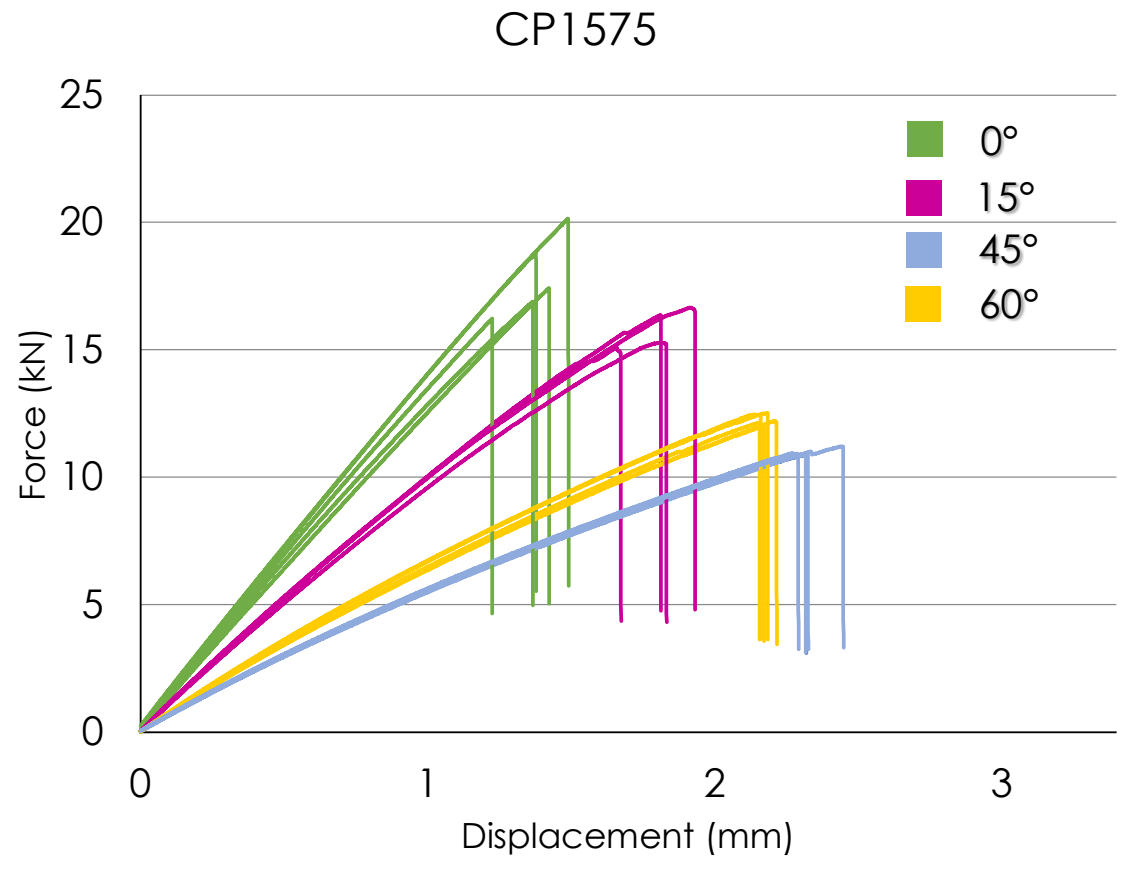

e)

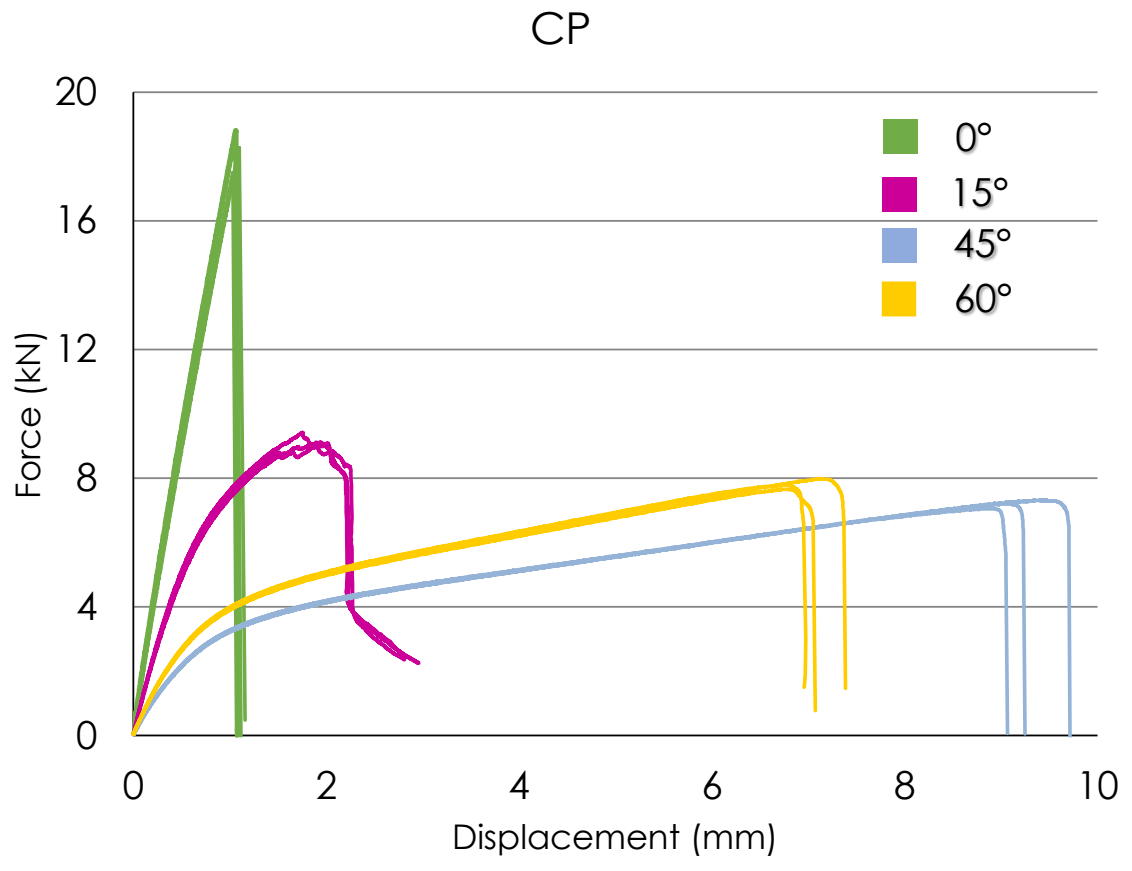

f)

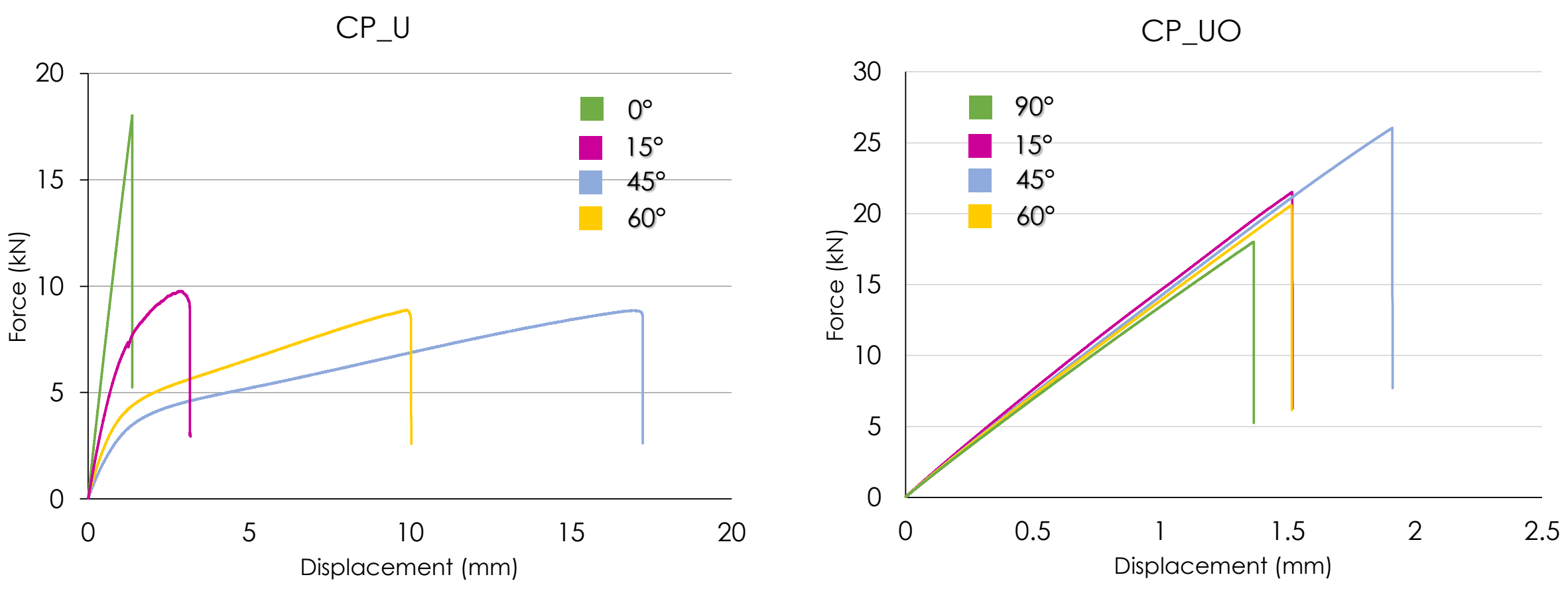

g) h)

**Fig. 2. Force-displacement curves for notched laminated composites under quasi-static loading** [22]. **Subfigures (a–h) represent configurations: a) QI, b) SOFT, c) CP30, d) CP1560, e) CP1575, f) CP with open-hole, g) CP with U-notch (CP_U) under off-axis loading (0°, 15°, 45°, 60°), and h) CP with U-notch under on-axis loading (CP_UO, notch angles 15°, 45°, 60°, 90°). Each colour denotes a specific off-axis or notch angle.**

From Fig. 2, it can be seen that the configurations exhibited a variety of failure behaviours, ranging from brittle to ductile, with a wide range of failure loads—from approximately 8 kN to 26 kN—and displacements at failure from less than 1.5 mm to more than 17 mm. This confirms the diversity and complexity of the experimental data available for validating the model.

Moreover, the diversity in observed failure modes is evident from the actual appearance of the failed samples and failure surfaces. Fig. 3 demonstrates representative failure patterns from six different notched specimens, highlighting some distinct failure modes and underlying mechanisms associated with the failure of the tested laminates. As reported in [22] the failure patterns observed varied widely—from brittle net-section, through-thickness cracks characteristic of thin-ply laminates, to more complex modes involving delamination, matrix cracking, and fibre pull-out also observed in thin-ply systems. The authors directly attributed these differences to the level of anisotropy in the laminates. They observed that specimens with lower anisotropy tended to fail in a brittle manner, regardless of the loading angle. In contrast, highly anisotropic laminates, especially under off-axis loading, exhibited more complex failure patterns involving extensive subcritical damage before final fracture.

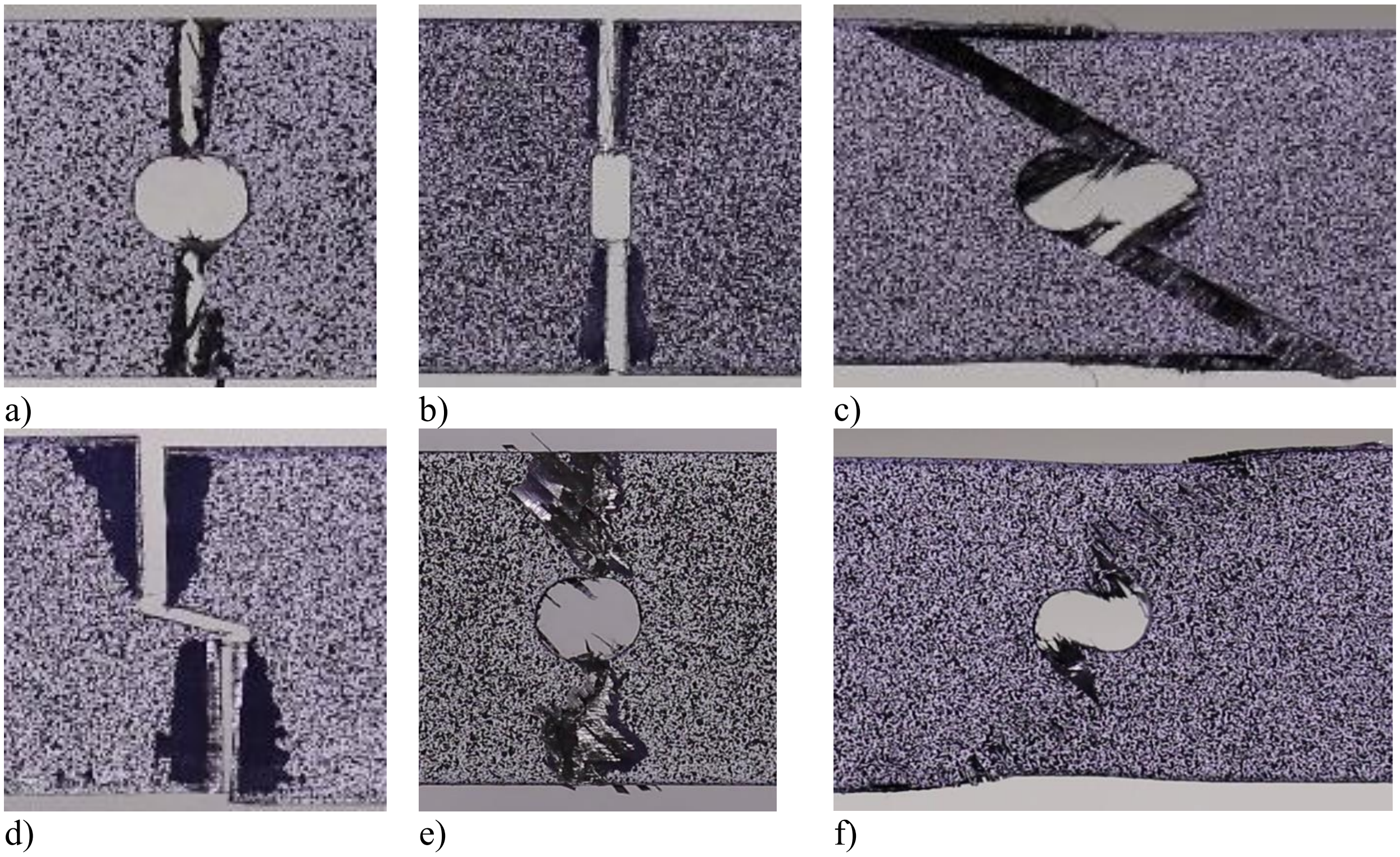

**Fig. 3. Representative post-failure fracture patterns in notched laminated composites, captured via optical photography** [22]**. Specimens include: a) SOFT_00 and b) CP_00_U (brittle, fibre-dominated net section failure), c) CP_60 (matrix-dominated shear-out with interlaminar shear), d) CP_15_UO (fibre-dominated net section failure along a path perpendicular to the load application direction), e) QI_45 (brittle, fibre-dominated net section failure), and f) CP_15 (localized band of rotated fibres accompanied by matrix damage).**

Overall, greater anisotropy in the laminate resulted in more pronounced changes in failure mode across different off-axis loading angles. The U-notched specimens exhibited similar failure mechanisms to the open-hole specimens; however, these failure processes were confined to a more localized region near the notch, Fig. 4. This increased localization is likely attributed to the higher stress concentration associated with the smaller notch tip radius. It is noted that the angle of the notch with respect to load direction, inducing a mixed mode state, appears to have minimal effect on the failure pattern since the crack still propagated from the strong concentration points in a net-section through thickness failure form along a path perpendicular to the load application direction. For more details and a more extensive discussion on the experimental findings refer to [22].

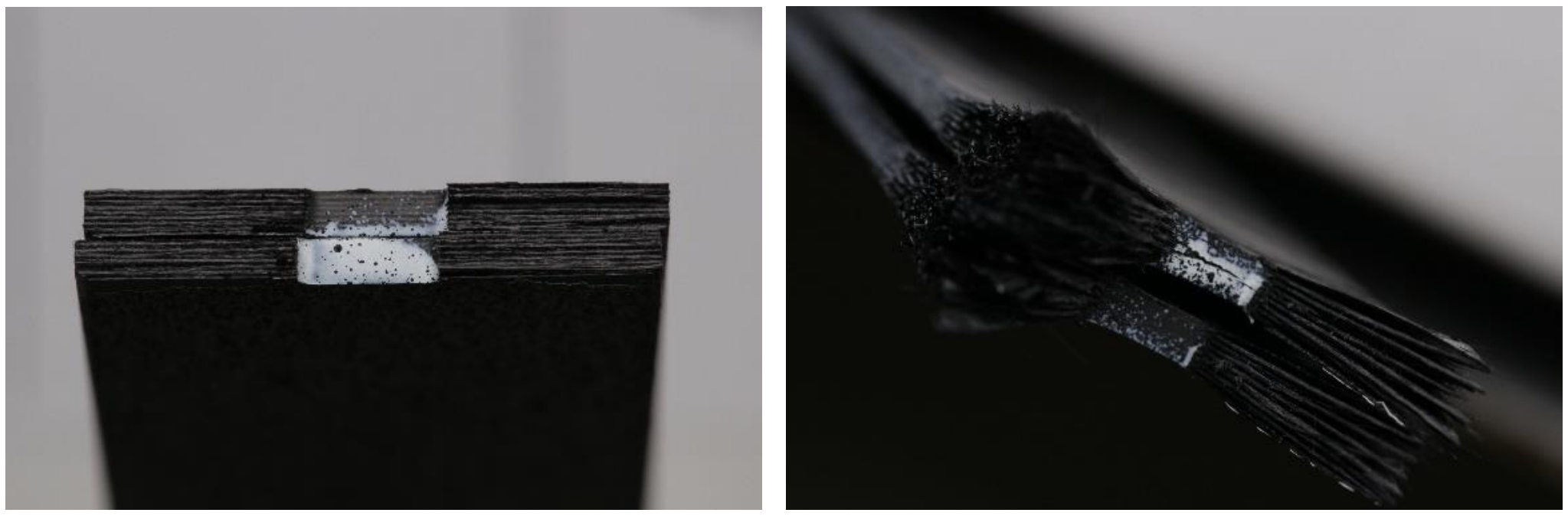

**Fig. 4. Representative post-failure fracture patterns in the U-notched laminated composites. Specimens include: a) CP_00_U (brittle, fibre-dominated net section failure), b) CP_45_U (fibre-dominated net section failure along a path perpendicular to the load application direction).**

## 3. Machine learning tools

### 3.1 Utilisation of DIC results in Deep Learning for fracture load prediction

The results obtained from DIC are utilized as inputs for machine learning (ML) algorithms to predict fracture loads. As mentioned earlier, here, the fracture load is defined as the maximum load that the specimen can withstand. Two distinct ML models were employed: the MLP and the CNN. The MLP model processes numerical values derived from DIC analysis, whereas the CNN model directly utilizes images of the strain field obtained from DIC. Specifically, the maximum principal strain field was selected as the indicator of failure at a predetermined load level across all specimens. Since the failure load is unknown, reference measures like 20% of the failure load, used in previous studies [15], cannot be employed to obtain a reference load to read the strain field. The chosen load level must be sufficiently high to minimize noise in the DIC measurements, yet low enough to stay below the minimum observed failure load so the model can be applied to every specimen. When no prior strength data of the sample are available, a simple physics-based rule of thumb calculations is useful. The laminate most likely to fail first is CP $[90/0]_{8S}$ cut at 45°, which behaves as a through-thickness [±45] laminate in axial tension. In this configuration, the response is shear-dominated (fibres are not aligned with the load), and first-ply failure is governed primarily by the in-plane shear strength rather than the fibre tensile strength. For an unnotched coupon, the axial capacity therefore scales with the ply shear strength and section size; introducing an open hole/notch reduces this capacity by a conservative notch-efficiency factor (<1). Under standard coupon geometries, even a conservative efficiency leaves a reasonable margin between this weakest-case and a reference load such as 7 kN. Therefore, 7 kN is a scientifically sound choice: it lies above the DIC noise floor yet below any possible failure load across all layups and off-axis angles, including the weakest and notched cases. The impact of varying this reference load level is further investigated in Section 4.2.

For the MLP approach, after computing the strain field at the selected load level, numerical values surrounding various notches were extracted. To address the variability in notch geometries, a simplified strategy was implemented: a rectangular region was defined ahead of the notch at a consistent relative distance from the notch tip, with fixed relative nodal positions. Key considerations include determining the rectangle's dimensions, the number of nodes it should contain, and its relative position to the notches. Previous studies by the author [19,21] suggest using as wide a region as possible; however, this is not mandatory, as evidence indicates that while nodes at greater distances contribute to predictions, their influence is less significant than that of nodes closer to the notch, where failure initiates and stresses are higher due to the notch presence. This will be further explored in Section 4.4. In simple terms, to balance computational efficiency, a relatively fine mesh, similar to that used in a proper finite element analysis of the specimen, is recommended to capture critical information around the notches without excessive computational cost. The effect of density of nodes in the region will be examined in Section 4.3. The rectangle should ideally be positioned as close as possible to the notch tip; however, due to reduced accuracy of DIC calculations (very) close to the notch boundary, proximity is limited. This limitation will be studied in Section 4.4. In this investigation, the rectangle spans from 0.1 mm away from the notch tip to 1 mm from the specimen edge, with the specimen edge located 7.5 mm from the notch tip (6.4 mm length). The longer side of the rectangle is twice the length of the shorter side and is symmetrically aligned with the notch bisector line. This area was discretized into a grid of 200 nodes (10 along the shorter side and 20 along the longer side), to consider the largest practically valid neighbourhood around the notch while avoiding boundary artifacts. The strain values at these nodes were flattened into a vector, serving as the sole feature for the MLP model. Note that for on-axis cases, the rectangle maintains the same relative position to the notch bisector line. The input dataset included this vector, specimen identifiers (used only for categorization to iterate over different specimens), and the corresponding failure loads. It is worth emphasizing that these identifiers are not included in the predictive input vector or used in model calculations and are only used to group test data.

For the CNN methodology [23,24], the strain field was extracted from a centered square region at the notch center, measuring 16 mm per side. To ensure data quality, a margin of 2 mm from the specimen's edge was excluded to eliminate the low accuracy and noise typically observed in DIC measurements near boundaries. These strain field maps were saved as images with a resolution of 224 × 224 pixels, a standard size in CNN due to its balance of feature capture and computational efficiency. To enhance prediction efficiency, the original three-channel (RGB) color images were converted to single-channel grayscale images. This transformation is appropriate since strain field data are scalar and can be effectively represented with a single channel, significantly reducing model complexity and data size. Prior studies confirm that reducing image channels lowers parameter count and computation time without compromising geometric information critical for model performance [25]. It is critical that the strain range remains consistent across all specimens, a condition achievable with many DIC software tools. The input data for the CNN analysis comprised these grayscale images and a file detailing the failure loads for each specimen.

### 3.2 Multi-Layer Perceptron (MLP)

This study develops a comprehensive and robust machine learning approach using an MLP neural network [26,27] to accurately predict fracture loads in composite materials. The method integrates advanced feature selection, cross-validation, hyperparameter tuning, and thorough training practices to ensure high accuracy predictions.

The input dataset comprises numerical features extracted from experimental setups and the related failure loads for each specimen. Given the complexity and noise in high-dimensional datasets, a detailed multi-step feature selection process is implemented. This begins with Mutual Information (MI) analysis [28], which evaluates the dependency between individual features and target variables to effectively identify the most relevant predictors. Mutual Information between a feature $X$ and the target variable $Y$ is defined as:

$$MI(X;Y) = \sum_{x \in X} \sum_{y \in Y} p(x,y) \log\left( \frac{p(x,y)}{p(x)p(y)} \right) \tag{1}$$

where $p(x, y)$ is the joint probability distribution, and $p(x)$ and $p(y)$ are the marginal probabilities. This step significantly reduces dimensionality by retaining approximately 90% of the most informative variables. The threshold of retaining 90% of the most informative variables after MI analysis was chosen to balance dimensionality reduction with information preservation. This cutoff was determined based on preliminary tests showing that lower thresholds led to a decrease in prediction accuracy, while higher thresholds increased computational cost with negligible gains. Following MI, the dataset undergoes further refinement using Lasso regression (Least Absolute Shrinkage and Selection Operator) [29] combined with cross-validation. Lasso regression minimizes the following objective function:

$$\min_{\beta}\left(\frac{1}{2n}\sum_{i=1}^{n}\left(y_i-\beta_0-\sum_{j=1}^{p}\beta_j x_{ij}\right)^2+\lambda\sum_{j=1}^{p}|\beta_j|\right) \tag{2}$$

where $n$ is the number of samples, $p$ is the number of features, $y_i$ is the target variable, $x_{ij}$ are the feature values, $\beta_j$ are the coefficients, and $\lambda$ is the regularization parameter. Lasso penalizes less relevant features by shrinking their coefficients towards zero, resulting in a sparse yet highly informative subset. The final stage of feature selection employs a Random Forest (RF) model [30,31] in conjunction with SHAP (SHapley Additive exPlanations) analysis [32,33]. After trial and error across preliminary tests, the final model employed the top 5 features to achieve the best combination of efficiency and accuracy.

SHAP values quantify the importance of each feature by calculating the average marginal contribution of a feature across all possible subsets of features:

$$\phi_j=\sum_{S\subseteq N\setminus j}\frac{|S|!(p-|S|-1)!}{p!}\left(f(S\cup j)-f(S)\right) \tag{3}$$

where $N$ is the set of all features, $S$ is a subset of features, $p$ is the total number of features, and $f(S)$ is the model output with features in $S$. This procedure yields a concise and reliable set of features tailored for accurate prediction.

To thoroughly evaluate the generalization capabilities of the model and prevent data leakage, the pipeline utilizes a two-layer cross-validation approach. An outer Leave-One-Group-Out (LOGO) cross-validation strategy is combined with an inner group k-Fold cross-validation [34] for hyperparameter optimization. The outer cross-validation loop was implemented with 31 folds, corresponding exactly to the 31 unique experimental configurations (combinations of layup, notch geometry, and loading angle), so that all specimens and repeats of that configuration are held out together, never mixing training and test data from the same geometry. Each outer fold isolates all specimens from a single configuration as the test set, ensuring that no data from the same physical scenario is seen during training or feature selection. Within each fold, scaling and feature selection are performed exclusively on the training subset to strictly prevent any possibility of data leakage. While this rigorous per-fold selection ensures no data leakage, a post-hoc analysis confirmed the stability of the process, revealing that a core subset of nodes—predominantly located in the physical region of highest stress concentration ahead of the notch—was consistently identified as most informative across the majority of cross-validation folds.

Within each outer fold, the inner loop implements group k-Fold splitting to accurately tune hyperparameters, ensuring results accurately reflect model performance rather than sample dependencies. Hyperparameters including hidden layer sizes, dropout rates and learning rates, are optimized using Optuna [35]—a Bayesian optimization framework. Optuna efficiently navigates the hyperparameter space through sequential model-based optimization, systematically searching for optimal values. Specifically, hidden layer sizes range from 32 to 128 neurons, dropout rates vary between 0 and 0.2, and learning rates span a logarithmic scale from 0.001 to 0.05. This strategic exploration provides comprehensive yet computationally efficient hyperparameter tuning.

The neural network architecture selected is a three-layer MLP [26,27] enhanced by batch normalization (BN) to stabilize and accelerate the training process. Hyperbolic tangent (Tanh) activation functions are utilized within each hidden layer, interspersed with dropout layers to prevent overfitting and enhance generalization. The employed MLP architecture can be represented as:

$$
\begin{aligned}
&\textbf{Input}: \mathbf{x} \in \mathbb{R}^d \\
&\textbf{HiddenLayers}: \textit{For} \quad i = 1,2,3: \\
&\qquad \mathbf{z}_i = \mathbf{W}_i \mathbf{h}'_{i-1} + \mathbf{b}_i \quad \text{with} \quad \mathbf{h}'_0 = \mathbf{x} \\
&\qquad \mathbf{z}'_i = \mathrm{BN}(\mathbf{z}_i) = \gamma_i \frac{\mathbf{z}_i - \mu_{\mathbf{z}_i}}{\sqrt{\sigma^2_{\mathbf{z}_i} + \epsilon}} + \beta_i \\
&\qquad \mathbf{h}_i = \tanh(\mathbf{z}'_i) \\
&\qquad \mathbf{h}'_i = \mathrm{Dropout}(\mathbf{h}_i, p_i) \\
&\textbf{OutputLayer}: \hat{y} = \mathbf{w}_4^\top \mathbf{h}_3 + b_4
\end{aligned} \tag{4}
$$

where $\boldsymbol{W}_i$ and $\boldsymbol{b}_i$ are the weights and biases, $p_i$ is the dropout probability, and $\hat{y}$ is the predicted fracture load. Training employs the Adam optimizer for efficient gradient descent. The Adam optimizer updates the parameters using:

$$
\begin{aligned}
&m_t = \beta_1 m_{t-1} + (1-\beta_1) g_t, \\
&v_t = \beta_2 v_{t-1} + (1-\beta_2)(g_t \odot g_t), \\
&\hat{m}_t = \frac{m_t}{1-\beta_1^t}, \quad \hat{v}_t = \frac{v_t}{1-\beta_2^t} \\
&\theta_t = \theta_{t-1} - \eta \frac{\hat{m}_t}{\sqrt{\hat{v}_t} + \epsilon}
\end{aligned} \tag{5}
$$

where $g_t$ is the gradient at time $t$, $\eta$ is the learning rate, $\beta_1$ and $\beta_2$ are exponential decay rates, and $\varepsilon$ is a small constant. Training efficiency is optimized further using batch-wise gradient updates and an early stopping mechanism based on validation loss performance. Additionally, an adaptive learning rate scheduler (ReduceLROnPlateau) [36] dynamically adjusts learning rates when training stagnation is detected, ensuring efficient convergence towards optimal model weights without unnecessary computational overhead.

Following hyperparameter optimization and training within each fold, comprehensive evaluation metrics—including Mean Absolute Error (MAE), Mean Absolute Percentage Error (MAPE), and the coefficient of determination ($R^2$)—are calculated. These metrics are defined as:

$$\mathrm{MAE} = \frac{1}{n}\sum_{i=1}^{n} |y_i - \hat{y}_i| \tag{6}$$

$$\mathrm{MAPE} = \frac{1}{n}\sum_{i=1}^{n} \left| \frac{y_i - \hat{y}_i}{y_i} \right| \times 100\% \tag{7}$$

$$R^2 = 1 - \frac{\sum_{i=1}^{n} (y_i - \hat{y}i)^2}{\sum i = 1^n (y_i - \bar{y})^2} \tag{8}$$

where $y_i$ is the true value, $\hat{y}_i$ is the predicted value, $\bar{y}$ is the mean of the true values, and $n$ is the number of samples.

These metrics provide detailed insights into predictive accuracy, error magnitude, and overall reliability. Results from all outer cross-validation folds are combined to deliver an extensive and holistic assessment of the model's predictive performance. Fig. 5 demonstrates the summary of the employed MLP algorithm for failure load prediction.

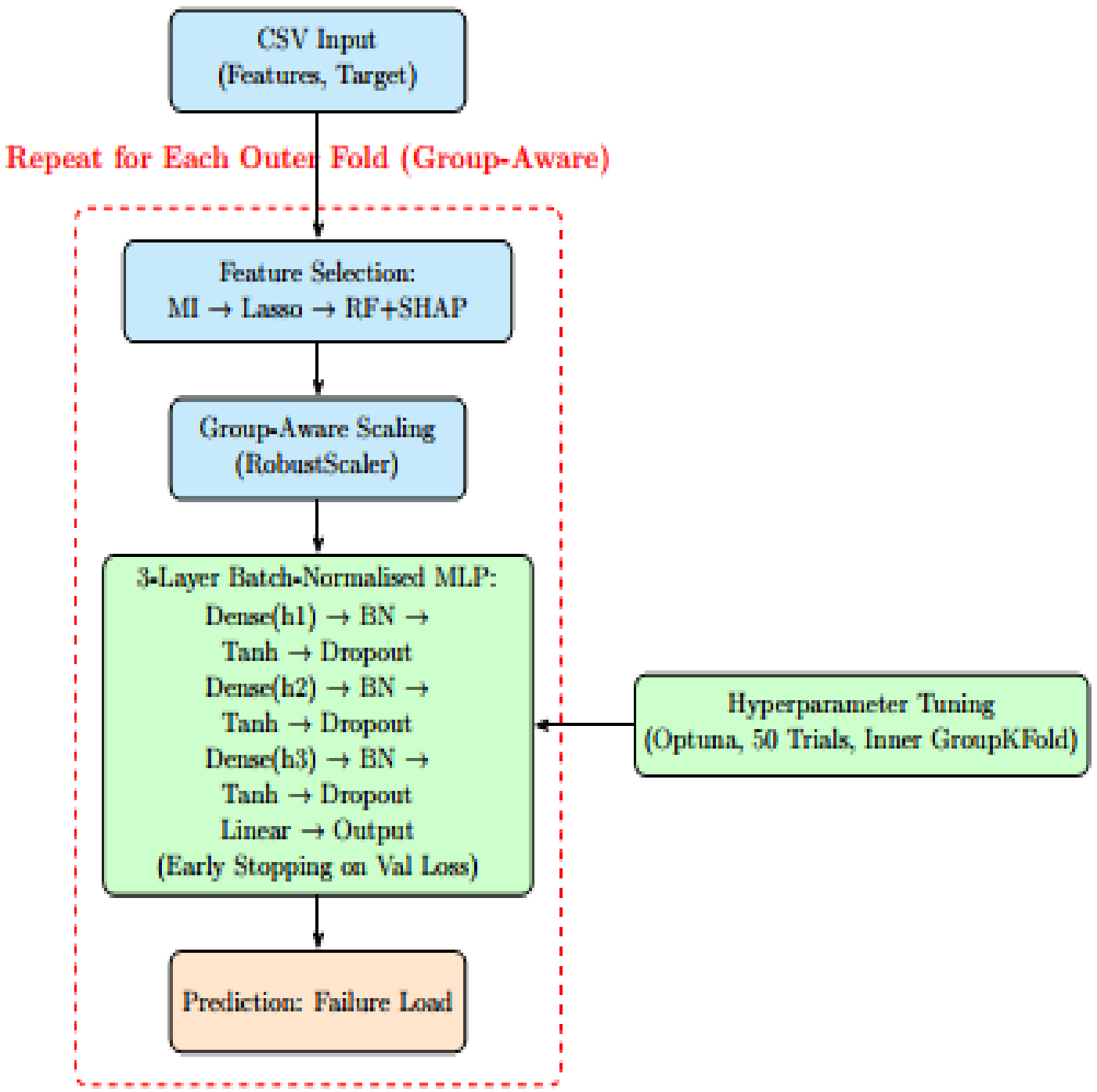

**Fig. 5. Schematic representation of the 3-layer batch-normalised MLP used for predicting fracture load from full-field numerical features derived from strain field data. The pipeline includes outer-loop cross-validation with group-wise holdout, followed by per-fold feature selection (mutual information, Lasso, and SHAP), robust group-aware scaling, and hyperparameter tuning via Optuna. The MLP architecture consists of three hidden layers, each followed by batch normalization, Tanh activation, and dropout. The final output layer predicts fracture load for each specimen.**

### 3.3 Convolutional Neural Network (CNN)

This study employs a CNN [23,24], specifically a modified DenseNet-121 [37] architecture, to predict fracture loads in notched laminated composites using strain field images. The developed method integrates image preprocessing, data augmentation, and training practices. These components are carefully designed to achieve high predictive accuracy, prevent data leakage, and ensure the generalizability and reliability of predictions. Given the sensitivity of fracture load predictions to strain variations near notches, precise and consistent image preprocessing is crucial. Fig. A.1 presents representative examples of strain field maps used as input for the CNN. The preprocessing pipeline features two primary enhancements: automatic notch masking via in-painting [38] and Contrast Limited Adaptive Histogram Equalization (CLAHE, clip limit = 2.0, tile grid = 8 × 8) [39]. Notch masking involves automatically detecting the notch region (black pixels) using adaptive Otsu thresholding (3 iterations with a 3 × 3 elliptical kernel) [40]. Otsu thresholding maximizes the inter-class variance:

$$\sigma_b^2(t) = w_0(t) w_1(t) (\mu_0(t) - \mu_1(t))^2 \tag{9}$$

where $w_0(t)$ and $w_1(t)$ are the probabilities of the two classes separated by threshold $t$, and $\mu_0(t)$ and $\mu_1(t)$ are their means. Morphological dilation [41] is then applied to refine the detected region. Dilation is defined as:

$$A \oplus B = z | \ (B)_z \cap A \neq \varnothing \tag{10}$$

This region is replaced with local mean grayscale values calculated from adjacent pixels, improving the signal-to-noise ratio by clearly revealing critical strain gradients around the notch. This in-painting strategy was specifically chosen over masking with a constant value to preserve the local continuity of the strain field data and prevent the introduction of artificial high-frequency edges at the mask boundary, which could be erroneously interpreted as physical features by the CNN's convolutional filters. Following notch masking, CLAHE is applied to enhance the contrast of subtle strain differences by equalizing local histograms within small image regions. This technique effectively highlighted key features necessary for the neural network's learning process. The chosen DenseNet-121 CNN architecture is pretrained on ImageNet to ensure robust initial feature extraction capabilities. Considering the images, the network's first convolutional layer was adapted to accept single-channel grayscale images rather than standard RGB channels. Because replacing the RGB stem re-initialises those weights, we freeze all remaining ImageNet-pre-trained layers and fine-tune the new 1-channel stem jointly with the classifier. Additionally, the final fully connected layer was modified to output a continuous scalar value predicting the fracture load.

To enhance model robustness and avoid overfitting, the training dataset undergoes data augmentation. This includes random horizontal and vertical flips, slight rotations (±10 degrees), and minor brightness and contrast adjustments [42]. It is worth emphasizing that these augmentations do not alter the underlying physics: they preserve the spatial distribution and magnitude of the principal-strain field around the notch (i.e., the salient gradient/intensity cues); this is consistent with Whitney–Nuismer model that failure is governed by the local field in a neighbourhood of the notch rather than its absolute orientation [3]. Each transform is applied on-the-fly with the default 0.5 probability in torchvision.transforms, yielding an average four-fold increase in distinct training samples per epoch. The modest rotation span and conservative color jitter widen the effective dataset while preserving the local strain-gradient patterns essential for accurate fracture-load regression. Small jitter helps robustness to minor DIC exposure/lighting changes. These augmentation techniques diversify the training data, allowing the CNN to learn generalizable features rather than memorizing specific pixel configurations. Note that all augmentations are applied only to the inner training subset within each outer fold; inner-validation and outer-test images are evaluated without augmentation.

The model optimization process involves a thorough two-stage training approach within each LOGO cross-validation fold, resulting in a total of 31 folds for 31 distinct test configurations. Initially, each training fold is internally partitioned into an 80% training and 20% validation subset, ensuring complete isolation of the test specimen to strictly prevent data leakage. It is worth emphasising that hyperparameter tuning is conducted exclusively on this internal subset, utilizing five chosen trials exploring critical parameters: learning rate (logarithmically sampled between $10^{-5}$ and $10^{-3}$), and weight decay (logarithmically sampled between $10^{-6}$ and $10^{-3}$). The AdamW optimizer is used and batch size is fixed at 8 to stabilise training on the relatively small dataset, with each trial trained for up to 50 epochs on the internal training set and evaluated on the internal validation set. This strategy is adopted to ensure a balance between accuracy and computational cost, as CNN-based analyses are considerably more computationally intensive than the MLP.

With optimized hyperparameters determined, the final CNN for each fold is trained on the complete internal dataset (combined training and validation subsets) for an additional 20 epochs. During this 20-epoch refit, no validation-driven decisions are made: early stopping and any validation-keyed schedulers are disabled. A fixed cosine-annealing schedule is used, and only the final checkpoint is evaluated on the outer-fold test set. The training utilizes the Smooth *L1* loss function [43] together with a very light target shrinkage ($\alpha = 0.01$) applied during training only to stabilise optimisation in small batches; validation and test use the original targets (no shrinkage). The Smooth *L1* loss is defined as:

$$\text{Smooth } L1(x) = \begin{cases} 0.5x^2 & |x| < 1 \\ |x| - 0.5 & \text{otherwise} \end{cases} \tag{11}$$

where $x = y - \hat{y}$, with $y$ the true label and $\hat{y}$ the predicted value. During training we replace each target by a lightly shrunk value toward the mini-batch mean:

$$y_i' = (1-\alpha) y_i + \alpha \bar{y}_B, \qquad \bar{y}_B = \frac{1}{|B|} \sum_{j \in B} y_j \tag{12}$$

where $\alpha$ is small shrinkage coefficient (set to 0.01), $\boldsymbol{B}$ denote the current training mini-batch and $\bar{y}_{\boldsymbol{B}}$ is the mini-batch mean target. This preserves the expected label mean and reduces batch-to-batch variance; no shrinkage is applied in validation or testing. A small ablation showed that ($\alpha$ = 0.01) lowered error slightly without altering conclusions, so we retain it.

Performance evaluation adhered strictly to the LOGO scheme, where the model predicts fracture loads for each unseen test configuration (31 in total) individually. Reproducibility and robustness are ensured by consistently setting model initialization seeds and data shuffling generators throughout all stages. Training dynamics are monitored using cosine annealing learning rate scheduling [44], facilitating stable convergence to optimal model parameters without manual intervention. Cosine annealing adjusts the learning rate as:

$$\eta_t = \eta_{\min} + \frac{1}{2}(\eta_{\max} - \eta_{\min})\left(1 + \cos\left(\frac{t}{T}\pi\right)\right) \tag{13}$$

where $\eta_t$ is the learning rate at epoch $t$, $\eta_{min}$ and $\eta_{max}$ are the minimum and maximum learning rates, and $T$ is the total number of epochs. We decay η from its initial value to $1 \times 10^{-6}$ using $T_{max} = 50$ in the search phase and $T_{max} = 20$ during the final 20-epoch fine-tune.

A schematic view of the developed CNN structure is shown in Fig. 6.

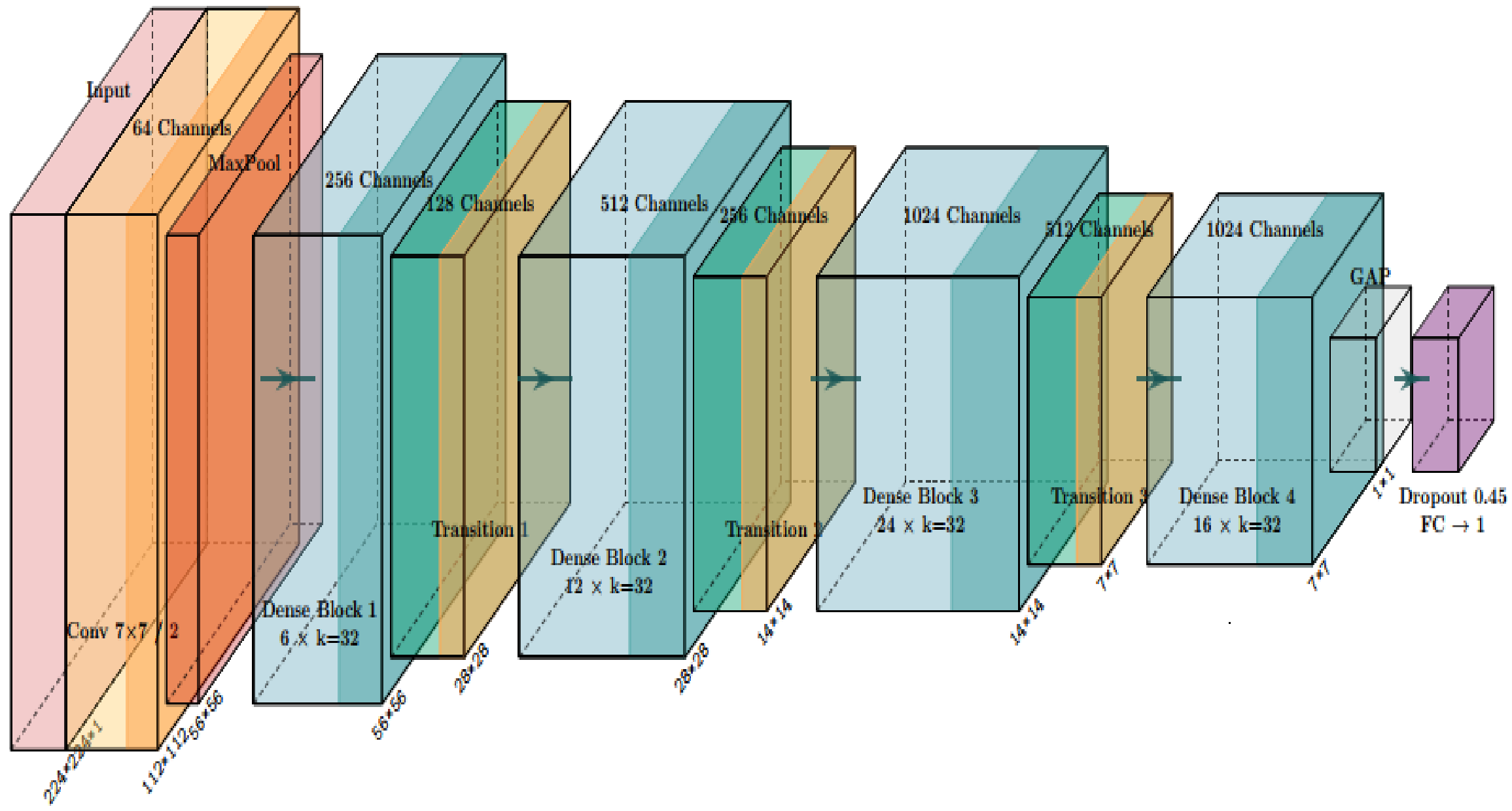

**Fig. 6. Schematic representation of the DenseNet-121-based CNN used for fracture load prediction from 224×224 grayscale strain field images. The network includes four Dense Blocks interleaved with Transition Layers. The final layers comprise a Global Average Pooling (GAP) and a fully connected regression output. The number of output channels and spatial dimensions are indicated for each stage.**

To justify the selection of DenseNet-121 for the CNN architecture, a comparative evaluation was conducted against other well-known models, including ResNet-50 and EfficientNet-B0, during the model selection phase. These models were chosen as candidates due to their established performance in image-based tasks and availability of pretrained weights on ImageNet, which facilitate robust feature extraction. However, ultimately, DenseNet-121 provided slightly more accurate predictions. The superior performance of DenseNet-121 might be attributed to its dense connectivity pattern, which promotes feature reuse and enhances the capture of complex strain gradients around notches, which is critical for accurate fracture load prediction in composites. Note that because we use only the maximum principal strain component, the MLP operates on a low-dimensional tabular input (after the three-step per-fold feature selection, reduces to five features), which reduces model complexity and the data needed for effective learning and generalisation. Also, the CNN ingests a single-channel image but still observes a high-dimensional field, so explicit regularisation (dropout, weight decay, early stopping) and physics-admissible augmentation remain essential to prevent overfitting.

## 4. Results and discussion

### 4.1 Prediction results

In this section, the fracture load predictions by the proposed framework for the notched laminated composites are presented. In order to compare the different cases and clearly demonstrate the accuracy of the framework, the analysis utilizes three key performance metrics: $R^2$, MAE, and MAPE. MAE reports the average magnitude of residuals in the same units as the response variable. $R^2$ expresses the proportion of observed variance explained by the model, while MAPE normalises residuals by the true values. To understand the accuracy of predictions across different specimens and loading conditions, the predictions by the MLP model are visualized in Fig. 7, while the predictions by the CNN model are shown in Fig. 8. The solid black line represents the ideal prediction, serving as a benchmark for comparison. Furthermore, ±20% scatter bands, defined as 0.8 and 1.2 times the ideal prediction, are depicted using red dashed lines to provide a clearer understanding of the model's performance and its deviation from perfect agreement with the experimental data.

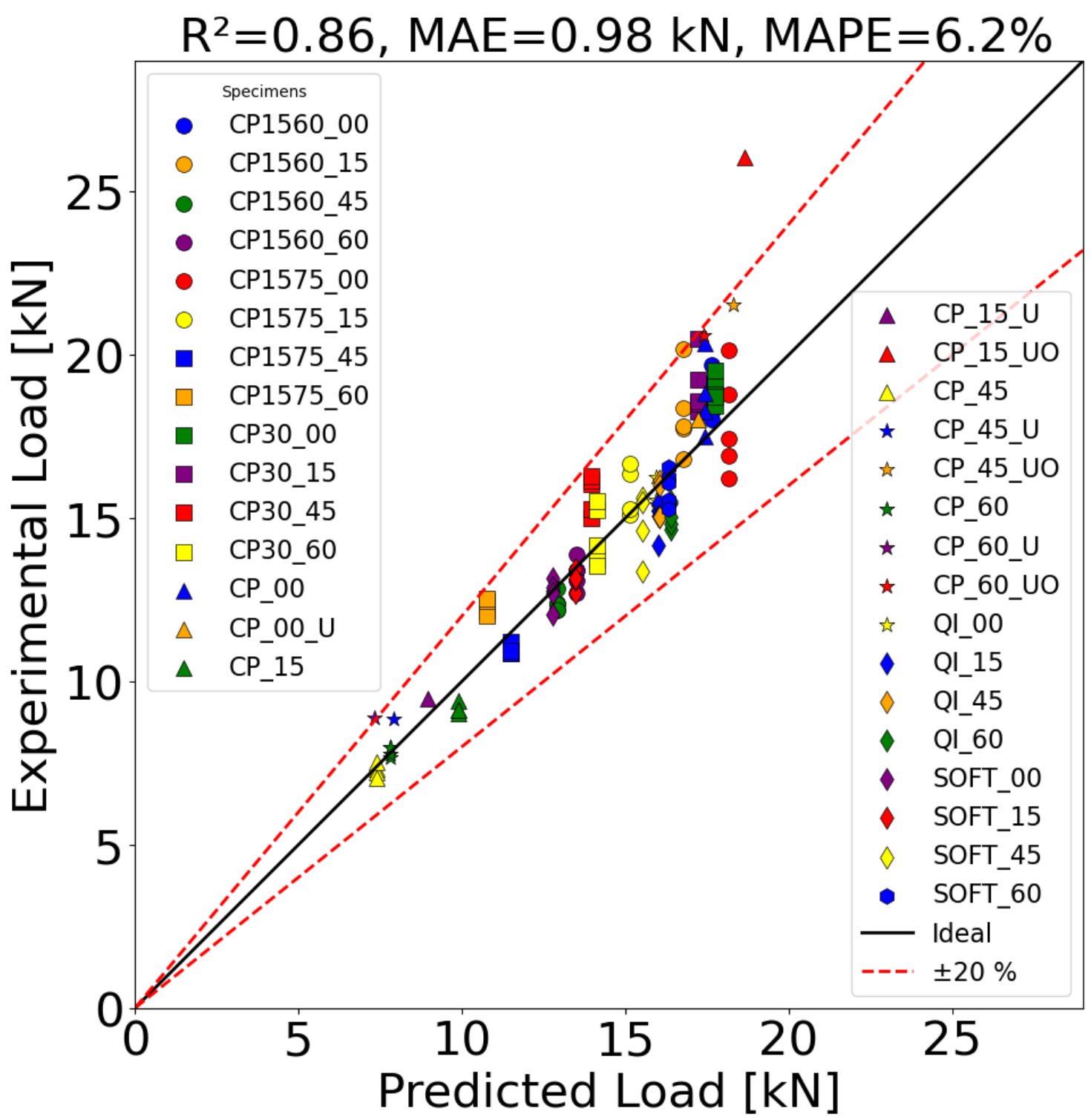

**Fig. 7. Predictions of fracture load for different specimens using the maximum principal strain field as input features in the MLP model. The solid black line represents the ideal prediction, while the red dashed lines indicate the ±20% scatter bands. The experimental data were obtained from [22].**

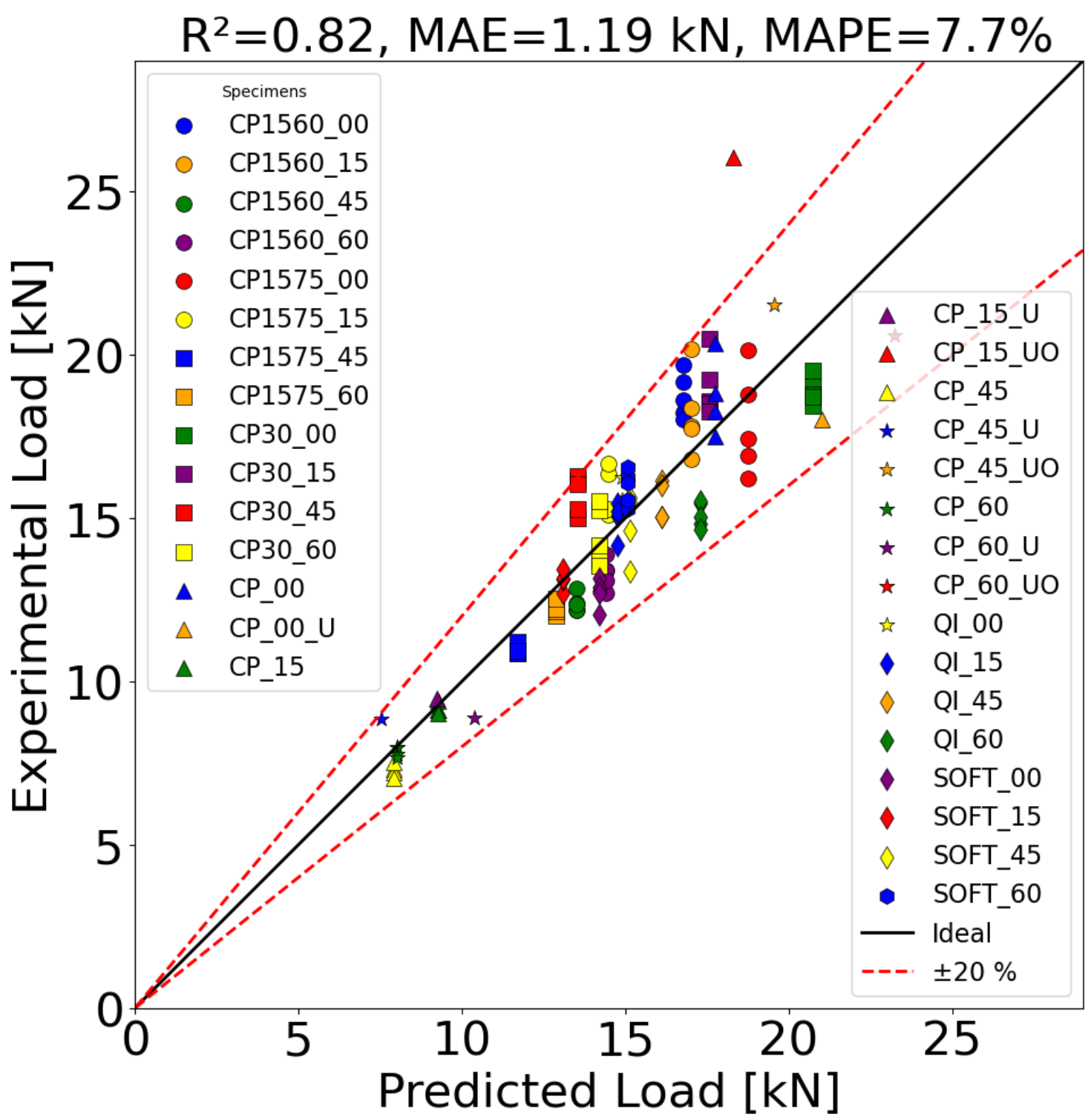

**Fig. 8. Predictions of fracture load for different specimens using the maximum principal strain field as input features in the CNN model. The solid black line represents the ideal prediction, while the red dashed lines indicate the ±20% scatter bands. The experimental data were obtained from [22].**

Considering Fig. 7 and Fig. 8, both models demonstrated excellent alignment with experimental data for fracture load predictions, with almost all predictions falling within the ±20% scatter bands, except for the specimen with a 15° U-notch. Focusing on Fig. 2 h), the failure loads for CP_15_UO and CP_60_UO are nearly identical, suggesting a possible issue with experimental testing which is also suggested by both models. The clustering of predictions for all different specimens with different notches, anisotropy levels, and off-axis angles around the ideal line suggests that the proposed framework effectively captures the mechanics driving fracture loads based on the principal strain field. In this case, the MLP achieved a slightly higher accuracy of $R^2$ = 0.86 compared to the CNN's $R^2$ = 0.82 for predicting fracture load using DIC data. Although CNNs are generally superior for spatial data due to their ability to detect local patterns, several factors likely contributed to the MLP's better performance. The main reason may be the limited size of the dataset. The employed CNN model contains around eight million parameters and thus needs heavy data augmentation, transfer learning, or weight decay to prevent overfitting. An MLP with three hidden layers and carefully tuned regularisation has a far lower number of parameters, so it can reach its bias–variance sweet spot sooner. On the other hand, MLP was fed by data only in the action zone. This focus reduces noise and maintains model accuracy. However, the CNN takes in the whole strain field image—everything from the critical notch area to far-off regions. It is worth noting that the results of the CNN model would decrease to $R^2$=0.57 without augmentation (see section 3.3), but still most data are within ±20% scatter bands. Additionally, the MLP's significantly lower computational cost (less than a few seconds on a standard computer once calibrated) makes it more suitable for real-time analysis in structural health monitoring applications. Overall, considering the complexity of the problem, it can be argued that both models provide remarkable predictive accuracy in an efficient way, without the need for any simulations, or further analyses.

### 4.2 Effect of load level selection

The load level at which strain field data are collected is a key factor in predicting fracture loads in this framework. Using DIC, the strain field must be captured at a load that provides clear, reliable data without causing any specimen to fail. Ideally, this load should be high enough to reduce noise in the measurements but low enough to stay below the smallest fracture load observed. Achieving this balance is essential for the deep learning models, MLP and CNN, to perform well. The main analysis used a load level of 7 kN, selected to be high enough to be able to represent the strain field correctly and exceed DIC noise thresholds while remaining below the minimum observed fracture load (~8 kN). At 5 kN, the same model parameters were used to test the same conditions under increased noise and different load level, with the results shown in Fig. 9 and Fig. 10. Note that the mean value of all failure loads is 14.722 kN, with 7 kN representing less than half of this value, and the maximum failure load is around 26 kN.

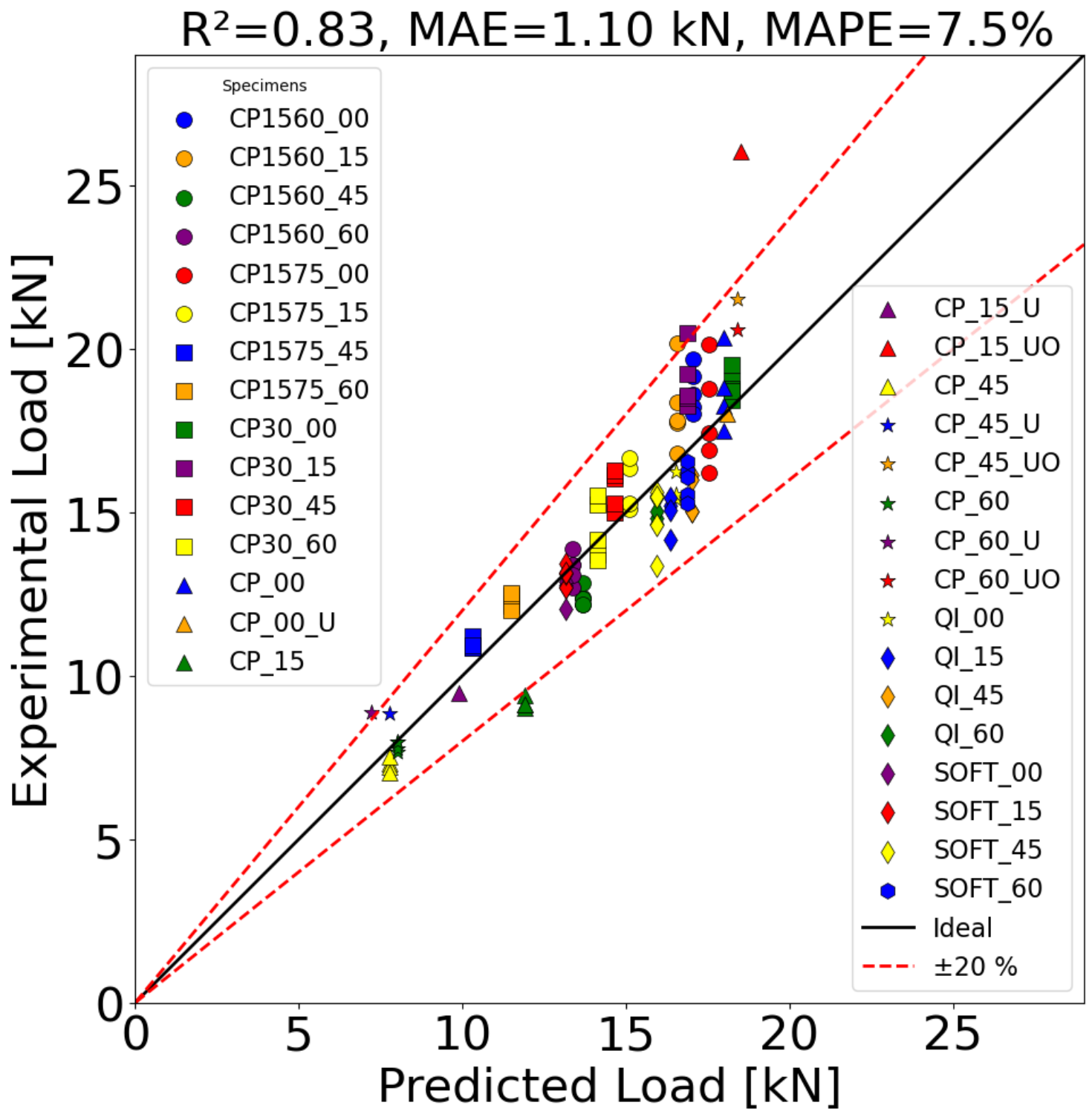

**Fig. 9. Predictions of fracture load for different specimens using the maximum principal strain field as input features in the MLP model evaluated at a low load level of 5 kN. The solid black line represents the ideal prediction, while the red dashed lines indicate the ±20% scatter bands. The experimental data were obtained from** [22].

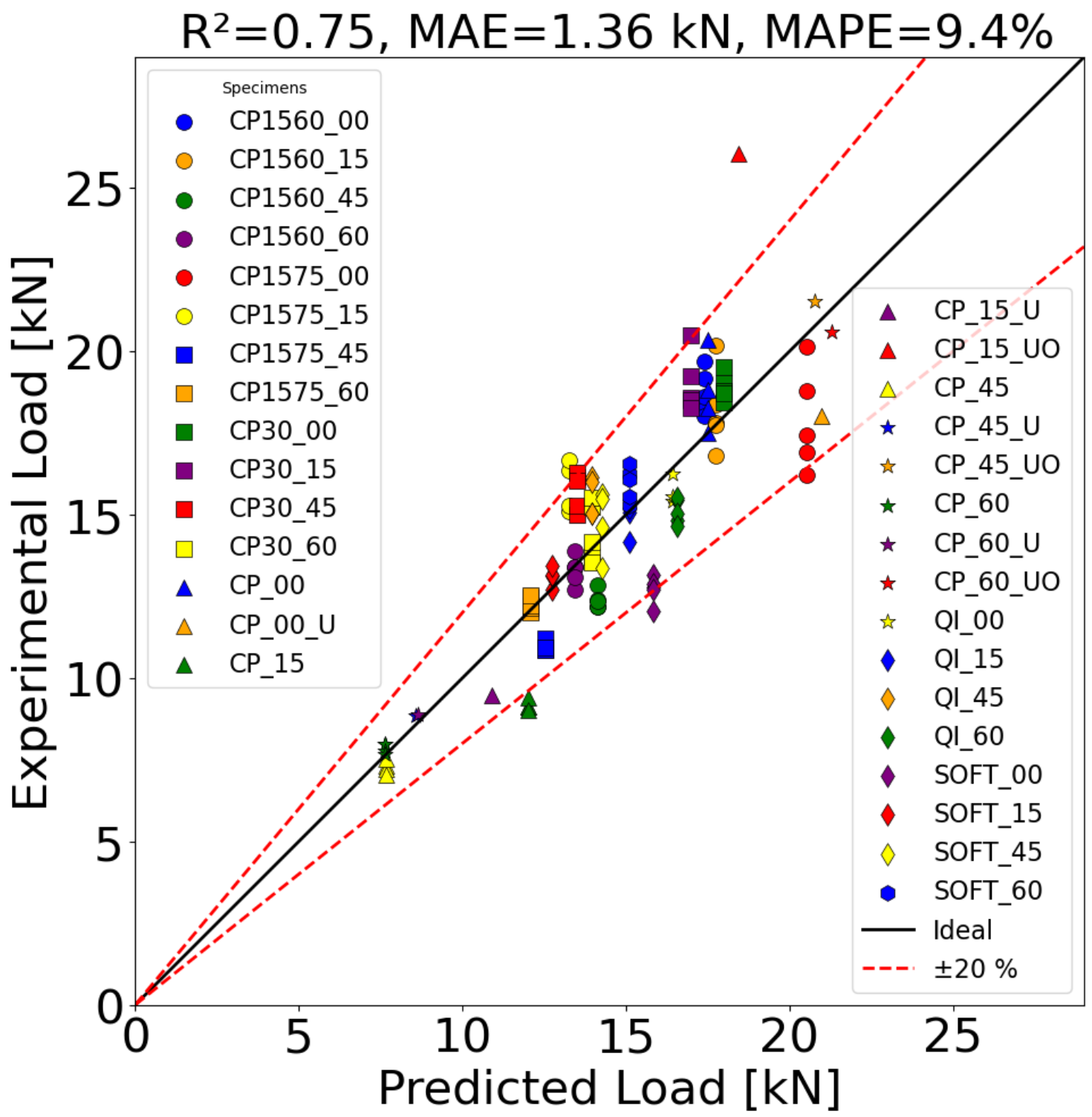

**Fig. 10. Predictions of fracture load for different specimens using the maximum principal strain field as input features in the CNN model evaluated at a low load level of 5 kN. The solid black line represents the ideal prediction, while the red dashed lines indicate the ±20% scatter bands. The experimental data were obtained from** [22]**.**

For the MLP model, Fig. 9 showed that predictions at 5 kN closely match those at 7 kN. Most data points stay near the ideal prediction line, which suggests the MLP handles lower load levels well, likely because it uses specific numerical features focused on strain near the notch, which remain clear even with some noise. The CNN model, however, performs differently. As seen in Fig. 10, its predictions at 5 kN scatter more widely around the ideal line compared to 7 kN. This drop in accuracy points to the CNN's sensitivity to noise in the strain field images at lower loads. Since the CNN processes full images (including strain patterns and minor details like speckle or lighting flaws), a lower load weakens the signal-to-noise ratio, making it harder for the model to focus on the most important features. It is worth emphasizing that the same parameters as 7 kN were used in both models. To investigate whether CNN performance at 5 kN could be enhanced, retuning hyperparameters such as the learning rate could potentially reduce sensitivity to lower signal-to-noise ratios in strain field images, which warrants further exploration in future studies.

### 4.3 Effect of number of data points

For the MLP model, an important consideration in implementing the framework is selection of representative nodes for the field distribution. The baseline analysis used a grid of 200 nodes (10 along the shorter side and 20 along the longer side of the rectangle) to extract strain values. To test the model's adaptability, two additional configurations were analysed: Case 1 with a high-density grid of 800 nodes (20 × 40), and Case 2 with a low-density grid of 50 nodes (5 × 10). These results are presented in Fig. 11 a) and Fig. 11 b) respectively. By comparing these setups, the analysis highlights how parameter selection influences the prediction accuracy of the proposed approach.

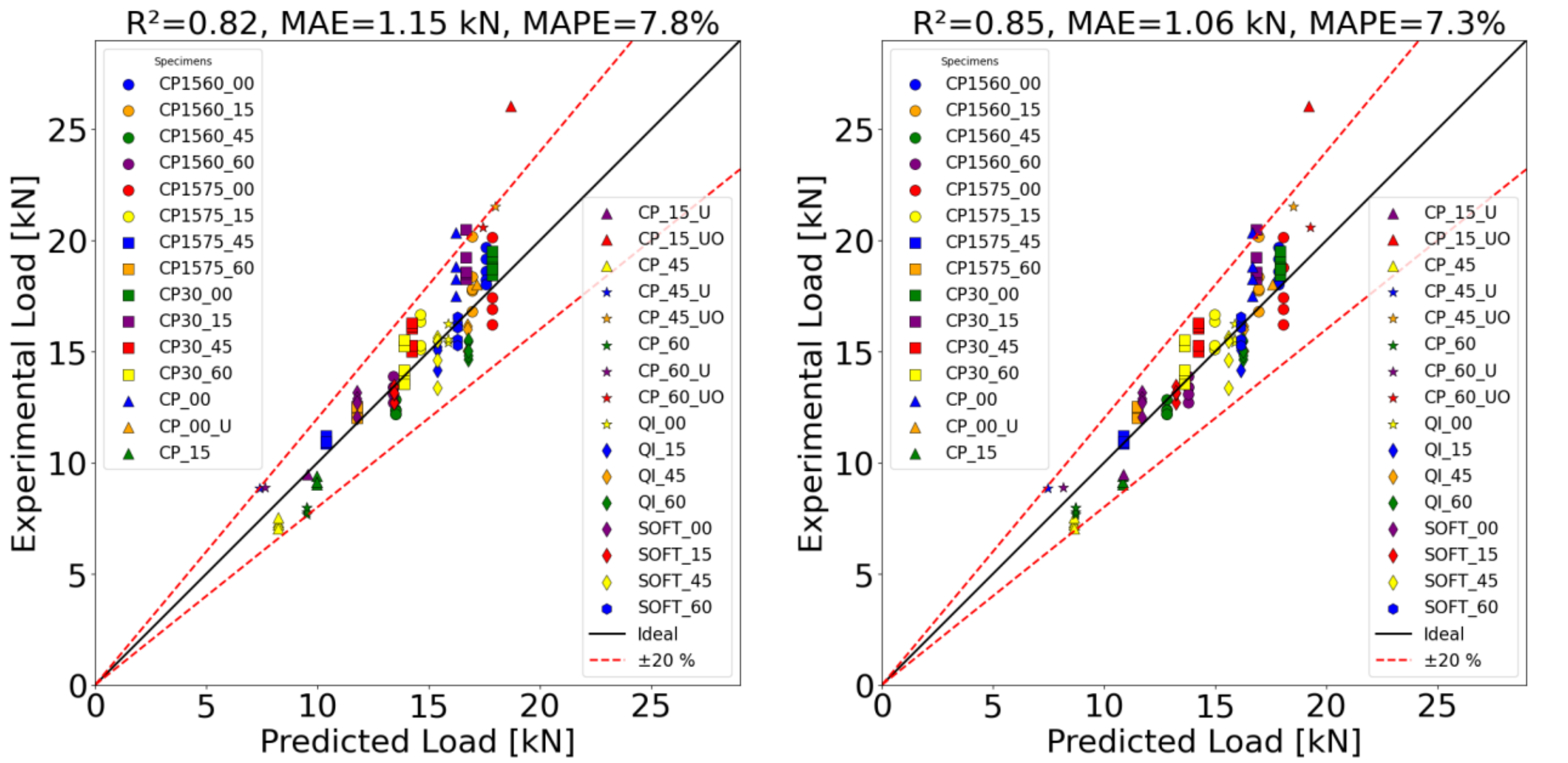

a) b)

**Fig. 11. Predictions of fracture load for different specimens using the maximum principal strain field as input features in the MLP model. The results are calculated based on discretizing the region of interest using a) 800**

**nodes, b) 50 nodes. The solid black line represents the ideal prediction, while the red dashed lines indicate ±20% scatter bands. The experimental data were obtained from** [22]**.**

Fig. 11 showed that for both high- and low-density configurations, predictions remained robust for most specimens. However, slightly larger deviations were observed for CP_15 and CP_45, which fall further outside the ±20% bands compared to the baseline. These findings demonstrate the flexibility of the feature selection algorithm in handling different node density levels, from 50 to 800 nodes, without significant loss of performance. The feature selection process plays an important role by identifying the most informative nodes and enhancing the efficiency of the framework. For practical applications, a moderate node density considering a typical finite element mesh for the problem (e.g., 200 nodes) offers an optimal balance between capturing essential strain data and minimizing computational cost.

### 4.4 Effect of closeness to the notch boundary

As mentioned, the MLP model relies on strain values extracted from a rectangular region positioned close to the notch. However, placing the sampling zone too close to the notch boundary risks unreliable measurements due to DIC limitations, such as edge effects or speckle pattern distortions. Positioning it too far may miss critical strain gradients. This section investigates how the distance of the rectangular region from the notch tip as well as its size affects the MLP's predictive performance, offering insights into optimizing this critical parameter for robust and practical applications.

The baseline configuration positioned the nearest edge of the sampling zone 0.1 mm from the notch tip and 1 mm from the specimen edge, yielding a rectangle with dimensions of 6.4 × 12.8 mm. To investigate the impact of proximity, two alternative configurations are tested: Case 1, with the rectangle starting 0.5 mm from the notch tip (sampling zone size: 6 × 12 mm), and Case 2, with the rectangle starting 0.3 mm from the notch tip (sampling zone size: 6.2 × 12.4 mm). Note that in the baseline configuration, the close proximity increases the likelihood of missing strain data due to DIC inaccuracies. To reduce this, a simple nearest-neighbor interpolation algorithm can be used to estimate missing values. Results for these configurations are shown in Fig. 12, with subfigures a) and b) representing the 0.5 mm and 0.3 mm distances, respectively.

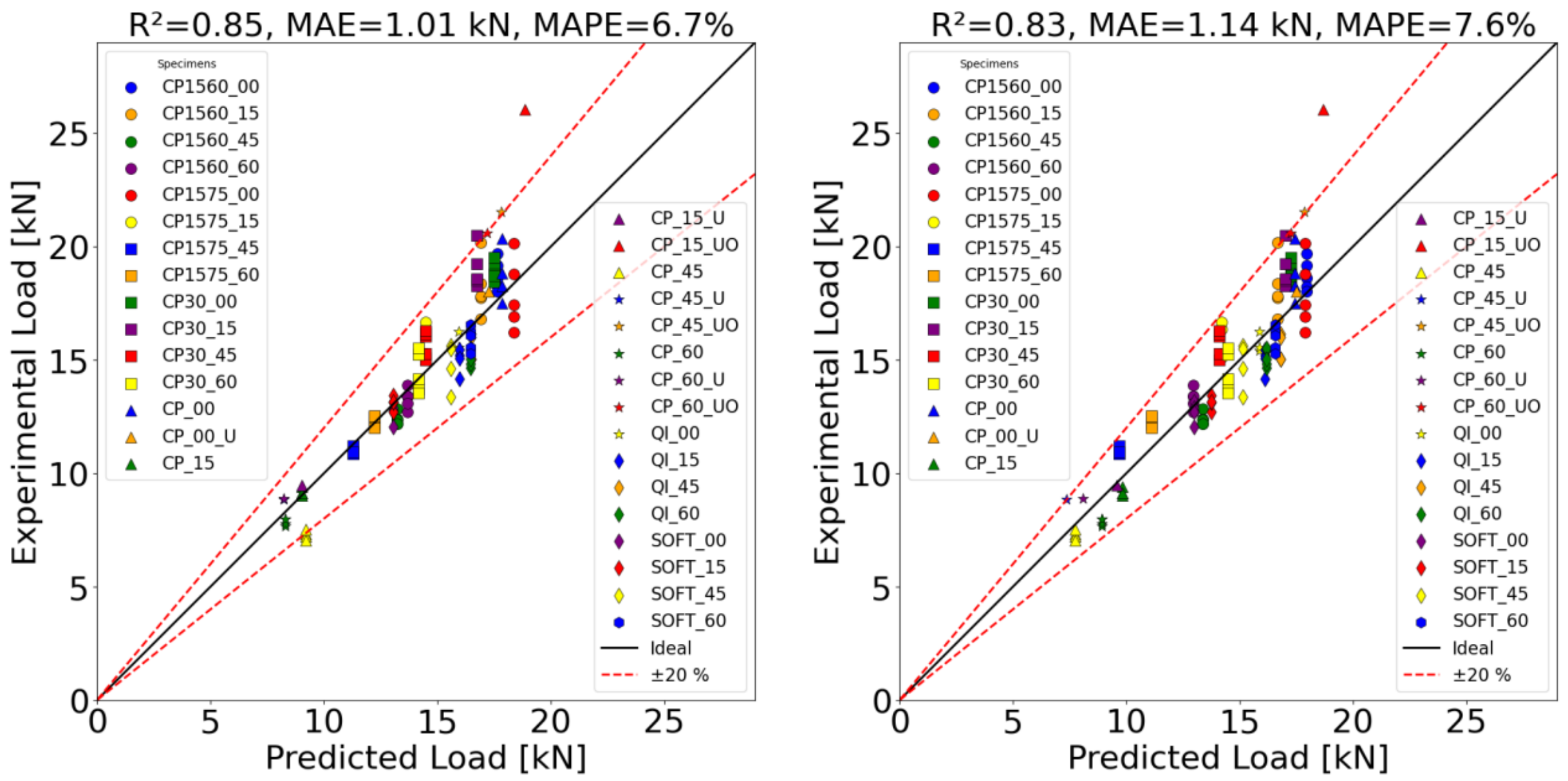

a) b)

**Fig. 12. Predictions of fracture load for different specimens using the maximum principal strain field as input features in the MLP model. The results are calculated for: a) 0.5 mm, and b) 0.3 mm distance between the region of interest and the notch tip. The solid black line represents the ideal prediction, while the red dashed lines indicate ±20% scatter bands. The experimental data were obtained from** [22]**.**

In Fig. 12 a), the 0.5 mm configuration produced predictions that closely follow the ideal line for most specimens, comparable to the baseline's performance (0.1 mm). Data points cluster tightly within the ±20% scatter bands, indicating that the MLP effectively captures essential strain patterns even at a slightly greater distance and smaller rectangle. Fig. 12 b) presents the closer distance to the notch tip, where predictions remain robust, with most points within the ±20% bands. The MLP's stable performance despite this challenge highlights the strength of the feature selection pipeline—utilizing Mutual Information, Lasso regression, and SHAP-based ranking—to prioritize high-impact strain values.

### 4.5 Discussion on feature selection algorithm

In this study, the framework predicts fracture loads based on strain values sampled from a grid of nodes near the notch. However, not all nodes contribute equally to the prediction; including too many nodes can introduce redundant information, increase computational cost, and potentially reduce model accuracy. Feature selection solves this by identifying the most critical nodes, ensuring the MLP focuses on the strain values that matter most. Fig. 13 provides insight into how features interact and contribute to the model's predictive performance. It should be noted that the results in this section are based on features from all groups, not one-leave-out algorithm. Also, in Fig. 13 b), the origin is at the centre of notches.

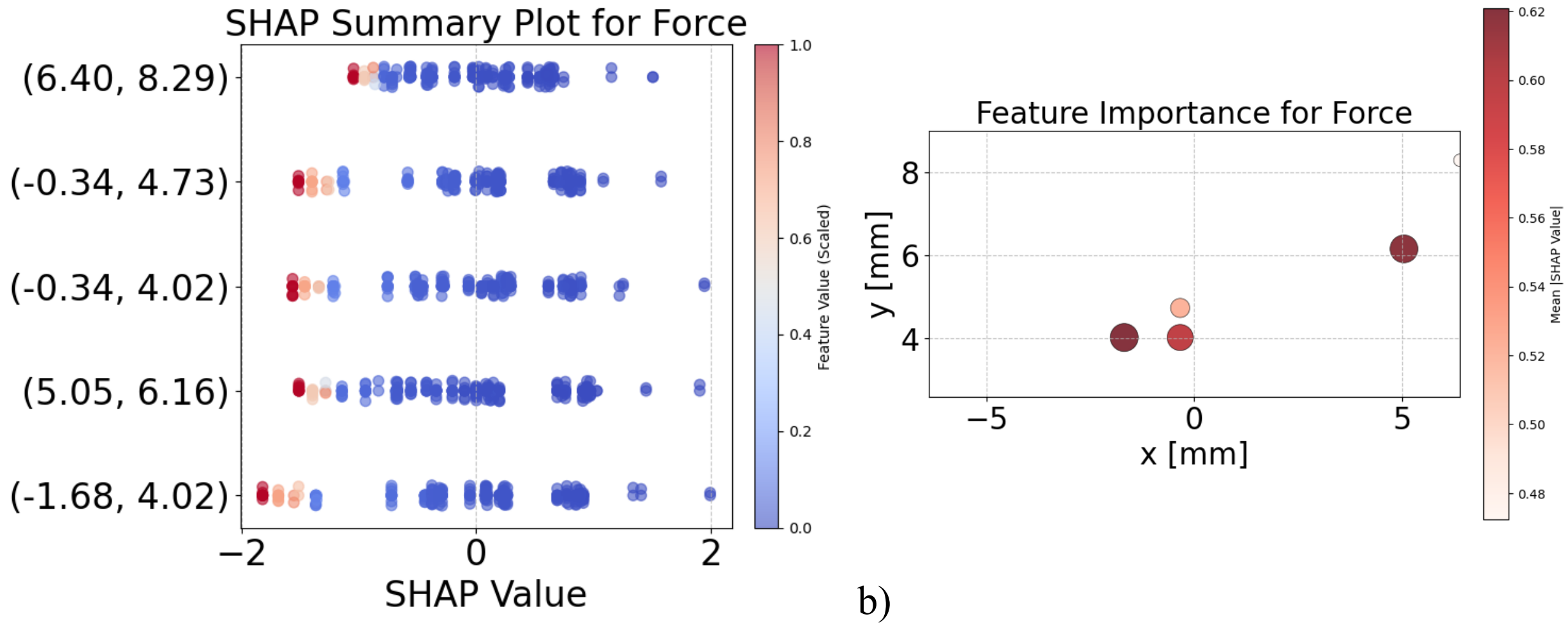

b)

a)

**Fig. 13. a) SHAP summary plot for the top 5 features presenting them in Cartesian coordinates. b) Feature importance based on SHAP value of top 5 features considering their locations. For the purpose of feature interpretation, this analysis was performed on a model trained on the full dataset.**

The SHAP summary plot in Fig. 13 a) illustrated the contribution of these five selected features to the predicted fracture load. Each point on the plot represented one data sample (i.e., one specimen test), with the x-axis showing the SHAP value, which quantifies how much that specific node's strain value influenced the model's output for that sample. Positive SHAP values indicated that the strain at that node increased the predicted force, while negative values indicated a decreasing effect. The color of each point represented the normalized strain value at that node—ranging from blue (low strain) to red (high strain). The two most influential nodes were located in close proximity to the notch tip, where local strain concentrations govern crack initiation. These nodes showed a wide range in SHAP values, signifying their dominant role in fracture load prediction. Notably, the most important node, at (-1.68, 4.02), contributed significantly to the model's predictions, where even relatively low strain values (blue points) are associated with high positive SHAP values—indicating that minimal deformation at this critical location is characteristic of higher load levels just before fracture initiation. In contrast, two nodes located in the far-field region (e.g., (6.4, 8.29)) exhibited negative SHAP values associated with high strain magnitudes. This pattern might reflect the physical behaviour of the specimen under increasing deformation—where strain accumulation in the far field is linked to anisotropy behaviour in material softening or crack propagation.

Following the analysis of the framework's behaviour under varying data selection strategies, the study acknowledges inherent limitations. The small dataset size limits both models, particularly the CNN's ability to fully exploit its spatial learning capabilities. Also, CNNs are computationally more expensive; this is the reason why further parameter tuning or analysis of the parameters' effect was not performed for the CNN model. Additionally, both models showed deviations for the 15° on-axis U-notch specimen, which may relate to variability in the experimental data.

Looking forward, several avenues can enhance this framework. Expanding the dataset with additional test cases, particularly for U-notch configurations and anisotropic layups, could improve the CNN's performance and generalization. Integrating hybrid models that combine the MLP's numerical precision with the CNN's spatial insights may yield even higher accuracy. Extending the framework to other loading conditions, such as compression or fatigue [45], would broaden its applicability in composite design and testing.

From a practical point of view, it can be argued that since the framework was able to successfully predict failure in notched laminated composites for a variety of anisotropy levels, loading conditions, and notch geometries using DIC, the previous problems solved by the same framework, like finite fatigue life of notched components across different materials [19], as well as fracture load and angle predictions in mixed mode loading conditions [21] can be tackled by the framework using DIC and in-situ online approaches.

## 5. Conclusions

This study presented a robust deep learning framework for in-situ prediction of fracture loads in notched laminated composites using only maximum principal strain field data acquired by DIC. The study employed two different deep learning approaches: an MLP that processes numerical strain values sampled from a grid of nodes near the notch, and a CNN that operates directly on full-field strain images extracted from DIC measurements to capture the complex interplay of notch geometry, fibre orientation, and layup. The experimental dataset used was notably complex, comprising 116 tests across 31 distinct configurations, including six laminate layups, four off-axis angles, and two notch geometries. This encompassed a wide range of anisotropy levels, both brittle and ductile behaviours, and diverse failure modes such as fibre-dominated fracture and delamination. The results demonstrated exceptional predictive accuracy, with the MLP achieving an $R^2 = 0.86$ and the CNN $R^2 = 0.82$, both exhibiting tight clustering of predictions within ±20% scatter bands. These results confirmed the framework's ability to model the intricate mechanics driving fracture in notched laminated composites.

The superior performance of the MLP highlighted the effectiveness of its targeted feature selection process, which combines Mutual Information, Lasso regression, and SHAP-based ranking to prioritize the most valuable information around the notch tip. This targeted, numerically-driven approach both enhanced accuracy and maintained computational efficiency, making the framework especially suited for real-time or resource-limited monitoring. In contrast, the CNN utilizes the detailed information in full-field DIC images to identify complex patterns, though its predictions are more sensitive to the selected load level. Sensitivity analyses further established that the MLP's predictive accuracy remains stable across a range of input node densities and sampling zone proximities, and that it is robust to moderate variation in DIC load levels, which is an important practical advantage. A key strength of this framework lies in its simplicity and practicality, as it requires no additional finite element analysis and allows for the direct use of full-field strain measurements. This streamlined approach is particularly valuable for in-situ structural health monitoring.

## Acknowledgment

The author expresses his sincere appreciation to Dr. Anatoli A. Mitrou for generously sharing raw videos of specimen deformation and the associated load–displacement data reported in [22], co-authored by Dr. Albertino Arteiro, Dr. José Reinoso, and Dr. Pedro P. Camanho.

## Appendix

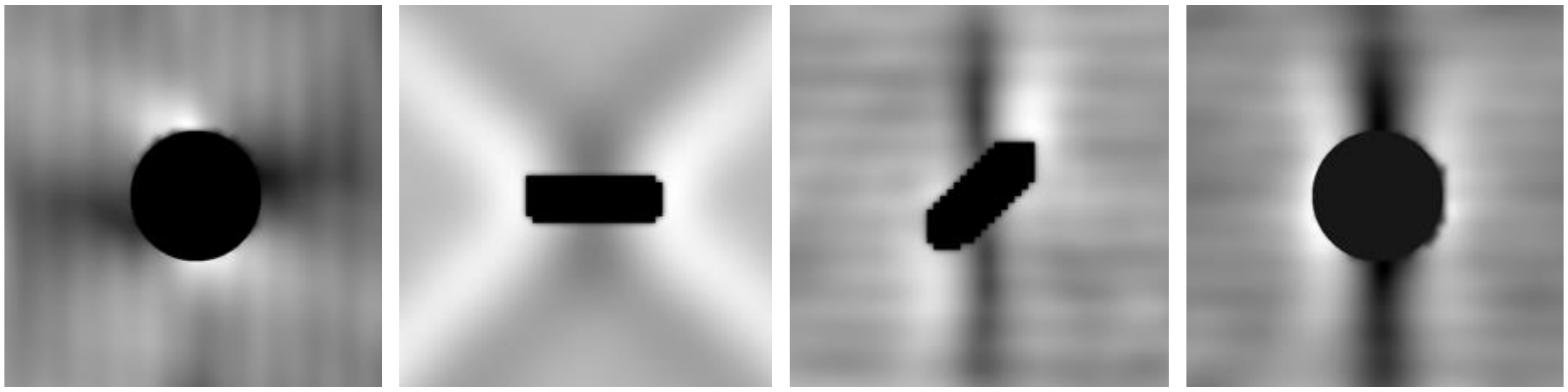

a) b) c) d)

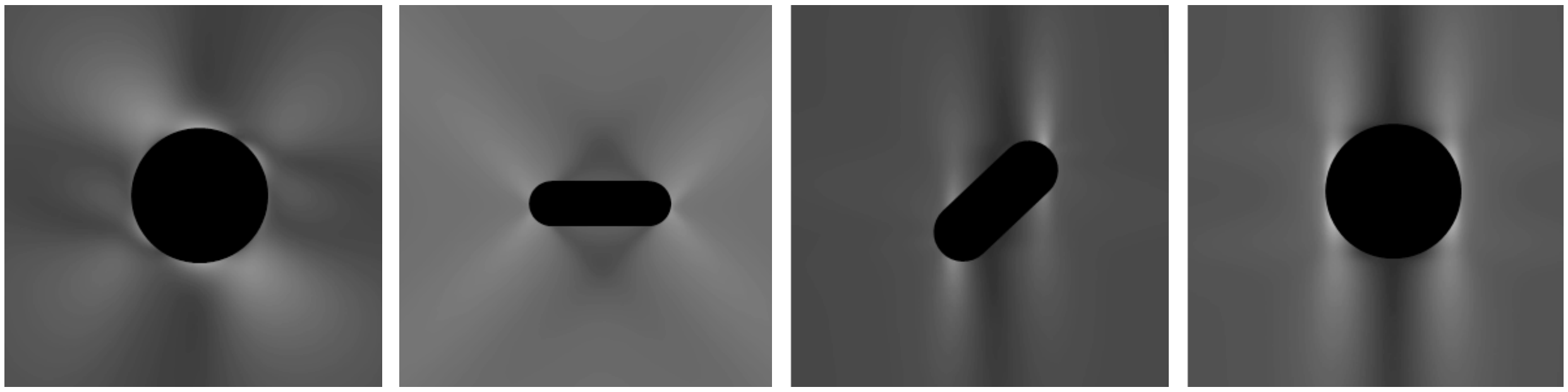

e) f) g) h)

**Fig. A.1. Maximum principal strain distributions for four composite specimens, showing experimental DIC measurements alongside finite-element validations. (a–d) DIC-measured strain fields for (a) CP1560_15, (b) CP_45_U, (c) CP_45_UO and (d) CP_00. (e–h) Corresponding FE predicted strain fields for (e) CP1560_15, (f) CP_45_U, (g) CP_45_UO and (h) CP_00. The FE contours are provided solely as a benchmark against the experimental DIC fields; no FE data were employed at any stage of the machine-learning model's training, validation, or inference.**